\documentclass[a4paper,12pt]{article}
\title{ {\LARGE\bf Conformal Field Theories of Stochastic Loewner
    Evolutions. }~~~~~~~~~~
[ CFTs of SLEs ]}
\date{}
\author{}

\newcommand{\mony}{f}
\newcommand{\monht}{ {\bf H}_t }

\usepackage[final]{graphicx}

\begin{document}
\maketitle

\vspace{-1.2cm}


\centerline{\large Michel Bauer\footnote[1]{Email:
    bauer@spht.saclay.cea.fr} and Denis Bernard\footnote[2]{Member of
    the CNRS; email: dbernard@spht.saclay.cea.fr}} 

\vspace{.3cm}

\centerline{\large Service de Physique Th\'eorique de Saclay}
\centerline{CEA/DSM/SPhT, Unit\'e de recherche associ\'ee au CNRS}
\centerline{CEA-Saclay, 91191 Gif-sur-Yvette, France}


\vspace{1.0 cm}

\begin{abstract}
  Stochastic Loewner evolutions $(SLE_\kappa)$ are random growth
  processes of sets, called hulls, embedded in the two dimensional
  upper half plane.  We elaborate and develop a relation between
  $SLE_\kappa$ evolutions and conformal field theories (CFT) which is
  based on a group theoretical formulation of $SLE_\kappa$ processes
  and on the identification of the proper hull boundary states. This
  allows us to define an infinite set of $SLE_\kappa$ zero modes, or
  martingales, whose existence is a consequence of the existence of a
  null vector in the appropriate Virasoro modules. This identification
  leads, for instance, to linear systems for generalized crossing
  probabilities whose coefficients are multipoint CFT correlation
  functions. It provides a direct link between conformal correlation
  functions and probabilities of stopping time events in $SLE_\kappa$
  evolutions. We point out a relation between $SLE_\kappa$ processes
  and two dimensional gravity and conjecture a reconstruction
  procedure of conformal field theories from $SLE_\kappa$ data.
\end{abstract}


\vskip 1.5 truecm

%
\def\tilde{\widetilde}
\def\bar{\overline}
\def\hat{\widehat}
\def\*{\star}
\def\({\left(}
\def\){\right)}
\def\zb{{\bar{z} }}
\def\frac#1#2{{#1 \over #2}}
\def\inv#1{{1 \over #1}}
\def\half{{1 \over 2}}
\def\d{\partial}
\def\vev#1{\langle #1 \rangle}
\def\bra#1{{\langle #1 |  }}
\def\ket#1{ | #1 \rangle}
\def\rvac{\hbox{$\vert 0\rangle$}}
\def\lvac{\hbox{$\langle 0 \vert $}}
\def\det{ {\rm det}}
\def\tr{{\rm tr}}
%
%
\def\th{\theta}         \def\Th{\Theta}
\def\ga{\gamma}         \def\Ga{\Gamma}
\def\be{\beta}
\def\al{\alpha}
\def\ep{\epsilon}
\def\la{\lambda}        \def\La{\Lambda}
\def\de{\delta}         \def\De{\Delta}
\def\om{\omega}         \def\Om{\Omega}
\def\sig{\sigma}        \def\Sig{\Sigma}
\def\vphi{\varphi}
%
\def\CA{{\cal A}}       \def\CB{{\cal B}}       \def\CC{{\cal C}}
\def\CD{{\cal D}}       \def\CE{{\cal E}}       \def\CF{{\cal F}}
\def\CG{{\cal G}}       \def\CH{{\cal H}}       \def\CI{{\cal J}}
\def\CJ{{\cal J}}       \def\CK{{\cal K}}       \def\CL{{\cal L}}
\def\CM{{\cal M}}       \def\CN{{\cal N}}       \def\CO{{\cal O}}
\def\CP{{\cal P}}       \def\CQ{{\cal Q}}       \def\CR{{\cal R}}
\def\CS{{\cal S}}       \def\CT{{\cal T}}       \def\CU{{\cal U}}
\def\CV{{\cal V}}       \def\CW{{\cal W}}       \def\CX{{\cal X}}
\def\CY{{\cal Y}}       \def\CZ{{\cal Z}}
\def\debut{ \begin{eqnarray} }
\def\fin{ \end{eqnarray} }
\def\non{ \nonumber }
%

\section{Introduction.}
Two dimensional conformal field theories \cite{bpz} have produced an
enormous amount of exact results for multifractal properties of
conformally invariant critical clusters. See eg.
refs.\cite{nienhuis,cardyconf,duplan} and references therein. These in
particular include the famous Cardy formula giving the probability
for the existence of a connected cluster percolating between two
opposite sides of a rectangle in two dimensional critical percolation
\cite{cardy2}.

This set of results triggered the search for a probabilistic, and more
rigourous, formulation of the statistical laws governing critical
clusters. It lead O. Schramm to introduce \cite{schramm0} the notion
of stochastic Loewner evolution $(SLE_\kappa)$.  These are conformally
covariant processes which describe the evolutions of random sets,
called the $SLE_\kappa$ hulls.  Two classes of stochastic Loewner
evolutions have been defined \cite{schramm0}: the radial and the
chordal $SLE_\kappa$.  They differ by the positions of the end points
connected by the random sets.  Below, we shall only consider the so-called
{\it chordal} $SLE_\kappa$.

The probabilistic approach has already lead to many new results,
including the Brownian intersection exponents obtained by Lawler,
Werner and Schramm \cite{LSW}.  It is also directly related to
Smirnov's proof of Cardy's formula and of the conformal invariance of
the scaling limit of critical percolation \cite{smirnov}.

Although motivated by results obtained using conformal field theory
(CFT) techniques, the relation between $SLE_\kappa$ evolutions and
conformal field theories in the sense of ref.\cite{bpz} remains
elusive and indirect. A step toward formulating this relation was done
in ref.\cite{bibi}. There we presented a group theoretical formulation
of $SLE_\kappa$ processes and we exhibited a relation between zero
modes of the $SLE_\kappa$ evolutions and conformal null vectors. The
aim of this paper is to develop this connection and to make it more
explicit.

Our approach is based on an algebraic -- and group theoretical --
formulation of $SLE_\kappa$ growths which takes into account the fact
that these processes are encoded in random conformal transformations
which are closely connected to special random vector fields defined
over ${\bf H}$.
This leads naturally to lift the $SLE_\kappa$ evolutions to Markov
processes in a group, whose Lie algebra is a Borel subalgebra of the
Virasoro algebra, and which may be identified with a Borel subgroup 
of the group of conformal transformations. 
By adjoint action on conformal fields, the
flows of these lifted processes induce that of the original
$SLE_\kappa$ evolutions.

The next key step consists in coupling the lifted $SLE_\kappa$
processes to boundary conformal field theories with specific Virasoro
central charges $c_\kappa<1$ depending on $\kappa$. These CFTs are
defined on random domains ${\bf H_t}$ which are the complement of the
$SLE_\kappa$ hulls in the upper half plane, see below for details.
Fluctuations and evolutions of the domains ${\bf H_t}$, or of the
$SLE_\kappa$ hulls, are then naturally encoded in boundary states,
called the hull boundary states, in the CFT representation spaces.
These states possess a natural geometrical interpretation in the
lattice statistical models underlying $SLE_\kappa$ evolutions.  We
show that these hull boundary states are zero modes of the
$SLE_\kappa$ evolutions, meaning that they are conserved in mean.  In
probabilistic terms, this also means that all components of these
states -- and there is an infinite number of them -- are local
martingales of the $SLE_\kappa$ evolutions.
 
This martingale property allows us to compute crossing probabilities
in purely algebraic terms. We show that, thanks to this property, the
generalized crossing probabilities are solutions of linear systems
whose coefficients are multipoint conformal correlations functions.
Stated differently, the martingale property of the hull boundary
states provides a direct link between $SLE_\kappa$ stopping time
problems and CFT conformal blocks.

Conformal field theories coupled to $SLE_\kappa$ hulls may be thought
of as CFTs in presence of 2D gravity since they are coupled to random
geometry. By analysing the behavior of the conformal correlation
functions close to the $SLE_\kappa$ hulls, this alternative viewpoint
leads us to an interesting connection between operator product
expansions in the presence of the $SLE_\kappa$ hulls and the KPZ
formula for gravitationally dressed scaling dimensions.  Relations
between scaling properties of critical clusters and 2D gravity are
crucial for the approach developed in ref.\cite{bertrand} and it would be
interesting to connect both approaches.

It is clear that the approach we present in this paper does not reach
the level of rigor of refs.\cite{LSW}. The aim of this paper is mainly to
show that a bridge linking the algebraic and the probabilistic
formulations of CFT may be developed.  As is well known, it is a
delicate matter to specify the appropriate group of conformal
transformations, and we do not give below a precise definition of the group
we use.  However, given the germs at infinity of the conformal transformations
$f_t(z)$, eq.(\ref{loew2}), which code for the $SLE_\kappa$ processes,
it is possible to give a constructive definition of the group elements,
which we shall call $G_t$, eq.(\ref{slevir}), when acting on highest weight
vector representations of the Virasoro algebra. We also do not specify
precisely the set of functions $F(G_t)$ for which the stochastic
equation eq.(\ref{itovir}) may be applied, although matrix elements of
$G_t$ in highest weight vector representations of the Virasoro algebra
clearly belong to this set. We would consider it an important research
project for mathematical physicists to make the connection between
$SLE_\kappa$ and CFT presented in this paper rigorous. For some work
in this direction see \cite{BBslevir}.

\vskip .5 truecm

The paper is organized as follows.  In Section 2, we recall the
definition of the stochastic Loewner evolutions and a few of its basic
properties.  We then present the group theoretical formulation of
these processes.  Section 3 is devoted to the algebraic construction
of $SLE_\kappa$ martingales, ie. zero modes of the $SLE_\kappa$
evolution operator, and to their link with null vectors in degenerate
Virasoro modules.  In Section 4, we give the geometrical
interpretation of these martingales and explain how they may be
identified with the hull boundary states. We also give there the
connection between CFTs coupled to $SLE_\kappa$ hulls and 2D gravity.
The algebraic derivation of generalized crossing probabilities is
described in Section 5. Section 6 gathers informations on the behavior
of conformal correlators when the hull swallows a domain. A conjectural
reconstruction scheme for conformal field theories based on the
relation between generalized crossing probabilities and multipoint CFT
correlation functions is proposed in the conclusions.

\vskip 1.0 truecm 

{\bf Acknowledgement:} We thank John Cardy, Philippe Di Francesco,
Antti Kupiainen, Vincent Pasquier and Jean-Bernard Zuber for
discussions and explanations on conformal field theories and
$SLE_\kappa$ processes.  This research is supported in part by the
European EC contract HPRN-CT-2002-00325.  \vskip 1.0 truecm

\section{Basics of SLE.}
The aim of this section is to recall basic properties of stochastic
Loewner evolutions (SLE) and its generalizations that we shall need in
the following.  Most results that we recall can be found in
\cite{schramm,lawler,LSW}. See \cite{cardynew} for a nice introduction
to SLE for physicists. 

\subsection{$SLE_\kappa$ processes.}
Stochastic Loewner evolutions $SLE_\kappa$ are growth processes 
defined via conformal maps which are solutions of Loewner's equation:
\begin{equation}
\partial_t g_t(z)=\frac{2}{g_t(z)-\xi_t}\ ,\quad g_{t=0}(z)=z
\label{loew}
\end{equation}
When $\xi_t$ is a smooth real-valued function, the map $g_t(z)$ is the
uniformizing map for a simply connected domain $\monht$ embedded in
the upper half plane ${\bf H}$, ${\rm Im}z>0$. At infinity $g_t(z)=z
+2t/z + \cdots $. For fixed $z$, $g_t(z)$ is well-defined up to the
time $\tau_z$ for which $g_{\tau_z}(z)=\xi_{\tau_z}$.  Following
refs.\cite{schramm,lawler}, define the sets $K_t=\{z\in{\bf H}:\ \tau_z\leq
t\}$.  They form an increasing sequence of sets, $K_{t'}\subset K_t$
for $t'<t$, and for smooth enough driving source $\xi_t$, they are
simple curves embedded in ${\bf H}$.  The domain $\monht$ is
${\bf H}\setminus K_t$.

$SLE_\kappa$ processes are defined \cite{schramm0} by choosing a
Brownian motion as driving term in the Loewner's equation:
$\xi_t=\sqrt{\kappa}\, B_t$ with $B_t$ a normalized Brownian motion
and $\kappa$ a real positive parameter so that ${\bf
E}[\xi_t\,\xi_s]=\kappa\,{\rm min}(t,s)$ \footnote{Here and in the
following, ${\bf E}[\cdots]$ denotes expectation (with respect to
the $SLE_\kappa$ measure), and ${\bf P}[\cdots]$ refers to
probability.}.  The growth processes are then that of the sets $K_t$
which are called the hulls of the process.
See refs. \cite{schramm, schramm0, LSW, lawler} for more details
concerning properties of the $SLE_\kappa$ evolutions.

It will be convenient to introduce the function 
$\mony_t(z)\equiv g_t(z)-\xi_t$
whose It\^o derivative is:
\debut
d\mony_t(z) = \frac{2}{\mony_t(z)}dt - d\xi_t 
\label{loew2}
\fin
Let $g^{-1}_t(z)$ and $\mony^{-1}_t(z)$ be the inverse of $g_t(z)$ and
$\mony_t(z)$ respectively. One has
$g^{-1}_t(z)=\mony^{-1}_t(z-\xi_t)$. For $t>s$ both
$(\mony_t\circ\mony_s^{-1})(z)$ and $\mony_{t-s}(z)$ have identical
distributions. In a way similar to the two dimensional
Brownian motion, $a^{-1}\mony_{a^2t}(az)$ and $\mony_t(z)$ also
have identical distributions.

\begin{figure}[htbp]
  \begin{center}
    \includegraphics[width=0.9\textwidth]{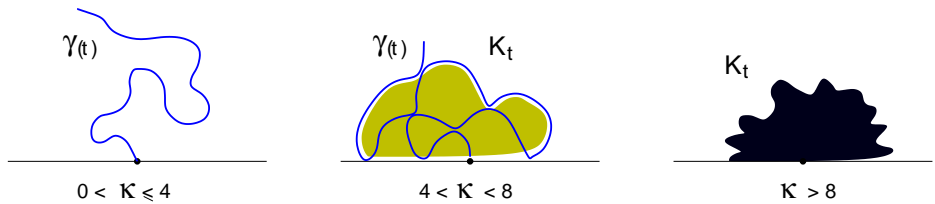}
      \caption{\em The different possible configurations of the
        $SLE_\kappa$ traces depending on $\kappa$.}
      \label{fig:sletrace}
  \end{center}
\end{figure}

The trace $\gamma[0,t]$ of $SLE_\kappa$ is defined by
$\gamma(t)=\lim_{\epsilon\to 0^+}\mony_t^{-1}(i\epsilon)$.  
Its basic properties were deciphered in
ref.\cite{schramm}. It is known that it is almost surely curve.
It is almost surely a simple (non-intersecting)
path for $0<\kappa\leq 4$ and it then coincides with the hull $K_t$;
for $4<\kappa<8$, it almost surely possesses double points but never
crosses itself, it goes back an infinite number of times to the real
axis and $\monht$ is then the unbounded component of ${\bf H}\setminus
\gamma[0,t]$; the trace of $SLE_\kappa$ is space filling for
$\kappa\geq 8$.  See Figure (\ref{fig:sletrace}).

The time $\tau_z$, which we shall refer to as the swallowing 
time of the point $z$, is such that  
$\mony_{\tau_z}(z)=0$. For $0<\kappa\leq 4$, the time $\tau_z$ is almost
surely infinite since the trace is a simple path,
while for $\kappa>4$ it is finite with probability one.
For $\kappa\neq 8$, the trace goes almost surely to infinity: 
$\lim_{t\to\infty}  |\gamma(t)|=\infty$.

A duality conjecture \cite{bertrand,schramm} states that the
boundary $\partial K_t$ of the hull $K_t$ of $SLE_\kappa$ with
$\kappa>4$ is statistically equivalent to the trace of the
$SLE_{\hat \kappa}$ process with dual parameter ${\hat \kappa}=16/\kappa<4$.

$SLE_\kappa$ evolutions may be defined on any simply connected domain
$\CD\subset {\bf H}$ by conformal transformations. Namely, let $a,b\in
\partial \CD$ be two points of the boundary of $\CD$ and $\varphi$ be
a conformal uniformizing transformation mapping $\CD$ onto ${\bf H}$
such that $\varphi(a)=0$ and $\varphi(b)=\infty$. Then the
$SLE_\kappa$ growth from $a$ to $b$ in $\CD$ is defined as that of the 
hull of the random map $g_t\circ \varphi$ from $\CD$ to ${\bf H}$. 

Two simple examples of determistic Loewner maps are presented in
Appendix A to help understanding the underlying geometry.

\subsection{Lifted $SLE_\kappa$ processes.} 
We now formulate the group theoretical presentation of $SLE_\kappa$
processes introduced ref.\cite{bibi}. It consists in viewing these
evolutions as Markov processes in the group of conformal
transformations.

Let $L_n$ be the generators of the Virasoro algebra $vir$ -- the 
central extension of the Lie algebra of conformal transformations -- 
with commutation relations
$$[L_n,L_m]=(n-m)L_{n+m}+\frac{c}{12}n(n^2-1)\delta_{n+m,0}$$
with $c$, the central charge, in the center of $vir$.  
Let $Vir\!_-$ be the formal group obtained by exponentiating the
generators $L_n$, $n<0$, of negative grade of the Virasoro algebra.
A possibly more rigorous  but less global definition of this group
consists in identifying it with the group of germs of conformal
tranformations $z\to z+\sum_{n>0} a_nz^{1-n}$ at infinity, the
group law being composition. 

We define a formal stochastic Markov process on $Vir\!_-$ by the first
order stochastic differential equation generalizing random walks on
Lie groups:
\begin{equation}
G_t^{-1}\, dG_t = -2dt\,L_{-2} + d\xi_t\, L_{-1}, \quad G_{t=0}=1
\label{slevir}
\end{equation}
The elements $G_t$ belong to the formal group $Vir\!_-$. 
We refer to the {\it formal} group $Vir\!_-$ because we do not specify
it precisely. We present in Section 3.3 simple computations
illustrating the relation between equations (\ref{slevir}) 
and (\ref{loew}) which are based on identifying $Vir\!_-$ 
with the group of germs of conformal
transformations at infinity. See also ref.\cite{BBslevir}.
Remark that for $t>s$, both $G^{-1}_sG_t$ and $G_{t-s}$ have 
identical distributions.

This definition is motivated by the fact that eq.(\ref{loew2}) may be
written as $\dot f_t(z)=2/f_t(z)-\dot\xi_t$ with $d\xi_t=\dot\xi_t
dt$. In this form eq.(\ref{loew2}) is slightly ill-defined, since 
$\dot\xi_t$ is not regular enough, but it nevertheless indicates that 
$SLE_\kappa$ describe flows for the vector field 
$\frac{2}{w}\partial_w-\dot{\xi_t}\partial_w$. 

Eq.(\ref{slevir}) is written using the Stratonovich convention for
stochastic differential calculus. This can be done using Ito
integrals as well.  Observables of the random process $G_t$ may be
thought of as functions $F(G_t)$ on $Vir\!_-$. Using standard rules of
stochastic calculus \footnote{We use formal rules extending those
valid in finite dimensional Lie groups, and we do not try to specify
the most general class of functions on which we may apply these
rules.}, 
one finds their It\^o
differentials: 
\debut d\, F(G_t) = {\cal A}\, \cdot F(G_t)\, dt + d\xi_t\,
\nabla\!_{-1} F(G_t)
\label{itovir}
\fin
with ${\cal A}$ the quadratic differential operator
\debut
\CA\equiv -2 \nabla\!_{-2} + \frac{\kappa}{2} \nabla\!_{-1}^{\ 2} 
\label{defcalA}
\fin
where $\nabla\!_n$ are the left invariant vector fields associated to the
elements $L_n$ in $vir$ defined by 
$(\nabla\!_n F)(G) = \frac{d}{du} F(G\, e^{uL_n})\vert_{u=0}$
for any appropriate function $F$ on $Vir\!_-$.
Since left invariant Lie derivatives form a representation of $vir$,
the evolution operator $\CA$ may also be written as:
\begin{equation}
{\cal A} = -2 L_{-2} +\frac{\kappa}{2}\, L_{-1}^2
\label{slehamil}
\end{equation}
with the Virasoro generators $L_n$ acting on the appropriate
representation space. 

In particular, the time evolution of expectation values of observables 
$F(G_t)$ reads:
\begin{equation}
\partial_t\, {\bf E}[F(G_t) ] = {\bf E}[\CA\cdot F(G_t) ]
\label{slevol}
\end{equation}

Notice that, in agreement with the scaling property of $SLE_\kappa$,
$\CA$ is homogenous of degree two if we assign degree $n$ to 
$\nabla\!_{-n}$.

\vskip .5 truecm

To test properties of the process $G_t$, we shall couple it to
a boundary conformal field theory (CFT) defined over $\monht$. We refer
to \cite{cardy} for an introduction to boundary CFTs that we shall mainly
deal with in the operator formalism -- see also Section 4. 
Its stress tensor,
$$
T(z)=\sum_n\, L_n\, z^{-n-2}\ , $$
is the generator of conformal
transformations and conformal (primary) operators are Virasoro
intertwiners.  We shall consider two kinds of conformal fields
depending whether they are localized on the boundary or in the bulk of
the domain.  More precisely, let $\Psi_\delta(x)$ be a boundary
intertwiner with dimension $\delta$. In our case the boundary of
$\monht$, away from the hull, is the real axis so that
$\Psi_\delta(x)$ depends on the real variable $x$.  By definition, it
is an operator mapping a Virasoro module $\CV_1$ into another,
possibly different, Virasoro module $\CV_2$ and satisfying 
the interwining relations: 
\debut [L_n,\Psi_\delta(x)]=\ell^\delta_n(x)\cdot
\Psi_\delta(x),\quad \ell_n^\delta(x) \equiv x^{n+1}\partial_x +
\delta(n+1) x^n. \label{inter1} 
\fin 
Bulk intertwiners $\Phi_{h,\bar
  h}(z,\bar z)$ with conformal dimensions $h$ and $\bar h$ are
localized in the bulk of $\monht$ and depend on the complex
coordinates $z$ and $\bar z$. They act from a Virasoro module to
another one and they satisfy the interwining relations: 
\debut
[L_n,\Phi_{h,\bar h}(z,\bar z)]=\(\ell^h_n(z)+\ell^{\bar h}_n(\bar
z)\)\cdot \Phi_{h,\bar h}(z,\bar z),\non\\
\quad ~~~~~~~~~ {\rm with}\quad \ell_n^h(z) \equiv z^{n+1}\partial_z +
h(n+1) z^n. \label{inter2} 
\fin

In the following, we shall only consider highest weight Virasoro modules.
More details on properties of interwiners, including in particular
the fusion rules they satisfy and the differential equations their
correlators obey, are briefly summarized in Appendix B. We shall need these
properties in the following Sections.

The above relations mean that
$\Psi_\delta(x)(dx)^\delta$ or $\Phi_{h,\bar h}(z,\bar z) (dz)^h(d\bar
z)^{\bar h}$ transform covariantly under conformal transformations.
As a consequence, the group $Vir$ acts on conformal primary
fields.  In particular for the flows (\ref{slevir}), one has: 
\debut
G^{-1}_t\, \Psi_\delta(x)\, G_t &=& [\mony_{t}'(x)]^\delta\,
\Psi_\delta(\mony_{t}(x)), \label{gonpsi}\\
G^{-1}_t\, \Phi_{h,\bar h}(z,\bar z)\, G_t &=& [\mony_{t}'(z)]^h\, [\bar
\mony_{t}'(\bar z)]^{\bar h}\,\Phi_{h,\bar
  h}(\mony_{t}(z),\bar\mony_{t}(\bar z)).  \non
\fin 
with $\mony_t(z)$ solution
of the Loewner's equation (\ref{loew2}) and
$\mony_{t}'(z) \equiv \partial_z\mony_{t}(z)$.  In words, by adjoint action
the flows of the lifted $SLE_\kappa$ implement the
conformal transformations of the $SLE_\kappa$ evolutions on the
primary fields.

Eqs.(\ref{gonpsi}) follow by construction but they may also be checked 
by taking their derivatives. Consider for instance a boundary primary field
$\Psi_\delta(x)$. By eq.(\ref{itovir}),
the It\^o derivative of the left hand side of eq.(\ref{gonpsi}) reads:
$$
d\({\rm l.h.s.}\)
= [2L_{-2}dt- d\xi_t L_{-1}, G_t^{-1}\Psi_\delta(x)G_t]+
\frac{\kappa}{2}[L_{-1}, [L_{-1},G_t^{-1}\Psi_\delta(x)G_t]]
$$
Similarly, the chain rule together with eq.(\ref{loew2}) and  the
intertwining relations (\ref{inter1}) gives for the It\^o derivative of 
the right hand side of (\ref{gonpsi}):
$$
d\({\rm r.h.s.}\)=[\mony_{t}'(x)]^{\delta}\{
[2L_{-2}dt-d \xi_t L_{-1},\Psi_{\delta}(\mony_{t}(x))]
+\frac{\kappa}{2}[L_{-1}, [L_{-1},\Psi_{\delta}(\mony_{t}(x))]]\}
$$
Making precise equations (\ref{gonpsi}) requires stating precisely
what is the domain of $G_t$, but eqs.(\ref{gonpsi}) clearly make sense
as long as the conformal fields are located in the definition domain
of the conformal map $f_t(z)$.

The occurrence of the Virasoro algebra in $SLE_\kappa$ evolutions may
also be seen by implementing perturbative computations valid close
to $z=\infty$, as explained below in Section 3.3.

\vskip 1.0 truecm

\section{Martingales and null vectors.}

The aim of this section is to show that martingales for $SLE_\kappa$
processes, which may be thought of as $SLE_\kappa$ observables which are
conserved in mean, are closely related to null vectors of appropriate
Verma modules of the Virasoro algebra.

\subsection{Modules and null vectors.}
Let us first recall a few basic facts concerning highest weight
modules of the Virasoro algebra \cite{bpz}. Such a module is generated
by the iterated action of negative grade Virasoro generators $L_n$, $
n \leq -1$ on a reference state, say $\ket{h}$, which is annihilated by
the positively graded Virasoro generators, $L_n\ket{h}=0$, $n>0$, and
has a given conformal weight $h$, $L_0\ket{h}=h\ket{h}$. These modules
are quotients of Verma modules. By definition, a Virasoro Verma
module, denoted $\CV_{h}$, is a highest weight module which is the
free linear span of vectors $\prod_pL_{-n_p}\ket{h}$ obtained by
acting with the $L_n$'s, $n<0$, on $\ket{h}$. Thus $\CV_{h}$ may be
identified with $(Vir\!_-)\ket{h}$. Generically Verma modules are
irreducible.

The Virasoro algebra acts on the space $\CV_h^*$
dual to the Verma module $\CV_h$ by $\bra{L_n\, v^*}=\bra{v^*}L_{-n}$
for any $v^*\in \CV_h^*$. The state $\bra{h}\in \CV_h^*$,
dual of the highest weight vector  $\ket{h}$ 
satisfies $\bra{h}L_n=0$ for any $n<0$  
and $\bra{h}L_0=h\,\bra{h}$. 

The vacuum state $\ket{0}$ is a highest weight vector with zero
conformal weight so that $L_n\ket{0}=0$ for all $n\geq-1$.
Its dual $\bra{0}$ satisfies $\bra{0}L_n=0$ for $n\leq 1$.

With $SLE_\kappa$ evolutions in view, we shall consider modules
with central charges 
\begin{equation}
c_\kappa = \frac{\(3\kappa-8\)\(6-\kappa\)}{2\kappa}
= 1 - \frac{3(4-\kappa)^2}{2\kappa}
\label{dingdong}
\end{equation}
It is then customary to parametrize the conformal weights as
\debut
h_{r,s}^{\kappa}=\frac{(r\kappa-4s)^2-(\kappa-4)^2}{16\kappa}
\label{hrs}
\fin
Notice that $c_\kappa=16\, h_{1,2}^{\kappa}h_{2,1}^{\kappa}<1$ and that
$h_{r,s}^{\kappa}=h_{s,r}^{\hat \kappa}$ with $\hat \kappa=16/\kappa$
the dual parameter.

For $r$ and $s$ positive integers, the corresponding Virasoro Verma module
$\CV_{h_{rs}^\kappa}$ possesses a highest weight vector submodule and is
reducible. This means that there exists a state 
$\ket{n_{r,s}}\in\CV_{h_{rs}^\kappa} $,
usually called a null vector, which is annihilated by the
$L_n$, $n>0$. If we assign degree $n$ to the $L_{-n}$, this state is
at degree $rs$. For generic value of $\kappa$, the quotient
$\CV_{h_{r,s}}/(Vir\!_-)\ket{n_{r,s}}$ is then an irreducible module, 
and we shall denote it by $\CH_{r,s}$.

\subsection{Zero modes and martingales.}

By definition, zero modes are observables which are
eigenvectors of the evolution operator ${\cal A}$ with zero 
eigenvalue so that their expectation is conserved in mean. 

We shall need details on the module $\CV_{h_{1,2}^\kappa}$ with weight 
$$
h_{1,2}^\kappa=\frac{6-\kappa}{2\kappa}
$$ 
We label its highest weight vector as $\ket{\om}$ 
so that $L_n\ket{\om}=0$ for
$n>0$ and $L_0\ket{\om}=h_{1,2}^\kappa\ket{\om}$. 
The null vector is at degree two and it is equal to
$\ket{n_{1,2}}=(-2L_{-2}+\frac{\kappa}{2}L_{-1}^2)\ket{\om}$. 
Indeed, let $\CA=-2L_{-2}+\frac{\kappa}{2}L_{-1}^2$ as in
eq.(\ref{slehamil}), then $\ket{n_{1,2}}=\CA\ket{\omega}$ and
$$ [L_n, {\cal A}] = (-2(n+2)+\frac{\kappa}{2}n(n+1))L_{n-2}
+ \kappa (n+1)L_{-1}L_{n-1} - c\delta_{n,2}
$$
Hence $L_n\ket{n_{1,2}}=[L_n,\CA]\ket{\om}=0$ for all $n>0$ since
$2\kappa h_{1,2}^\kappa=6-\kappa$ and
$c_\kappa=h_{1,2}^\kappa(3\kappa-8)$. The quotient
$\CH_{1,2}\equiv \CV_{h_{1,2}}/(Vir\!_-)\ket{n_{1,2}}$ is generically 
irreducible, and $\CA\ket{\om}=0$ as a vector in $\CH_{1,2}$.  

Let us now define the observable $F_\om(G_t)\equiv G_t\ket{\om}$ where
the group element $G_t$ is viewed as acting on the irreducible module
$\CH_{1,2}$ -- not on the reducible Verma module $\CV_{h_{1,2}}$. In
particular, $F_\om(G_t)$ is a vector in the infinite dimensional space
$\CH_{1,2}$.

By construction, $F_\om(G_t)$ is a zero mode. Indeed
$(\CA\cdot F_\om)(G_t)=
G_t\,(-2L_{-2}+\frac{\kappa}{2}L_{-1}^2)\ket{\om}=G_t\ket{n_{1,2}}$ 
vanishes  as a vector in $\CH_{1,2}$, since 
$\ket{n_{1,2}}=\CA\ket{\omega}$ does in $\CH_{1,2}$. 
By eq.(\ref{slevol}), the expectation ${\bf E}[F_\om(G_t)]$ is thus
conserved: 
\debut 
\partial_t\, {\bf E}[\, G_t \ket{\omega}\, ] =0 
\label{conserv}
\fin
The stationary property of the expectation of 
$F_\omega(G_t)$ is a direct consequence of the existence of a null 
vector in the corresponding Virasoro Verma module.  
\vskip .5 truecm

Summarizing leads to the following
\footnote{For simplicity, we shall write `martingales' instead of
the more appropriate denomination `local martingales'.}

{\bf Proposition.} {\it Let $\ket{\om}$ be the highest weight
vector of the irreducible Virasoro module with central charge
$c_\kappa= (3\kappa-8)(6-\kappa)/2\kappa$ and conformal weight
$h_{1,2}^\kappa=(6-\kappa)/2\kappa$. Then
$\{F_\om(G_t)\equiv G_t\ket{\om}\}_{t\geq 0}$ is
a martingale of the $SLE_\kappa$ evolution, meaning that for $t>s$:
\debut
{\bf E}[\, G_t\ket{\om}\, \vert \{G_{u\leq s}\}\,]= G_s\ket{\om}
\label{martin}
\fin
}

This is a direct consequence of the conservation law (\ref{conserv})
and of the fact that $G^{-1}_sG_t$ and $G_{t-s}$ are identically
distributed.

Using Dynkin's formula, see ref.\cite{oksen} p. 118,
the above proposition has the following

{\bf Corollary.} {\it Let $\tau$ be a stopping time
such that ${\bf E}[\tau]<\infty$, then:
\debut
{\bf E}[\, G_\tau\ket{\omega} \,] = \ket{\omega}
\label{stop}
\fin
}

All components of the vector ${\bf E}\,[\, G_t \ket{\omega}\, ]$ are
conserved and we may choose the vector on which we
project it at will depending on the problem.
The most convenient choices of vectors will be those generated by 
products of conformal operators. Namely,
\debut
\bra{\chi}\ \prod_j\Phi_{h_j,\bar h_j}(z_j,\bar z_j)\cdot
\prod_p\Psi_{\delta_p}(x_p) \non
\fin
with $\bra{\chi}$ the dual of a highest weight vector with
weight $h_\chi$. Since $\bra{\chi}L_n=0$ for any $n<0$, we have
$$\bra{\chi} G_t^{-1} =\bra{\chi}$$
because $G_t\in Vir\!_-$.
As a consequence, the conservation law (\ref{conserv}) projected on
these vectors  reads:
\debut
&& \partial_t {\bf E}[\
\bra{\chi}\ \prod_j [\mony_t'(z_j)]^{h_j}[\bar \mony_t'(\bar z_j)]^{\bar 
  h_j}\Phi_{h_j,\bar h_j}(\mony_t(z_j),\bar \mony_t(\bar z_j))\cdot \non \\
&& ~~~~~~~~~~ \cdot
\prod_p[\mony_t'(x_p)]^{\delta_p}\Psi_{\delta_p}(\mony_t(x_p))\ket{\om}\ ]=0
\label{bigbig}
\fin
where we moved $G_t$ to the left using the intertwining relations
(\ref{gonpsi}).   
\medskip

{\bf Example.} 
The simplest example is provided by considering correlations of the
stress tensor $T(z)$. For instance, from eq.(\ref{conserv}) we have:
$$
{\bf E}[\, \bra{\omega} T(z) G_t\ket{\omega}\, ]
 = \bra{\omega} T(z) \ket{\omega}
$$
For non-vanishing central charge, $T(z)$ transforms anomalously under
conformal transformations \cite{bpz} so that:
$$
G^{-1}_t T(z) G_t = [\mony_t'(z)]^2\, T(\mony_t(z)) + \frac{c_\kappa}{12}
\{\mony_t(z),z\} 
$$
with $\{\mony_t(z),z\}$ the Schwarzian derivative of $\mony_t(z)$.
Since $\bra{\omega} T(z)\ket{\omega}=z^{-2}\, h_{1,2}^\kappa$
and $\bra{\omega}G_t^{-1}T(z)\ket{\omega}=\bra{\omega}T(z)\ket{\omega}$, 
we get:
$$
{\bf E}[\ h_{1,2}^\kappa \Big({\frac{\mony_t'(z)}{\mony_t(z)}}\Big)^2 + 
\frac{c_\kappa}{12}\{\mony_t(z),z\}\ ]=  \frac{h_{1,2}^\kappa}{z^2}
$$
This may be checked by a perturbative expansion in $1/z$ around
$z=\infty$.
The extension to an arbitrary number of stress tensor insertions
is straightforward.

\subsection{Perturbative computations in SLE.}

The group $Vir$ acts on germs of 
meromorphic functions with a pole at infinity
$z(1+a_1/z+a_2/z^2+\cdots)$ with $a_k$ as coordinates, similar to
Fock space coordinates. The Virasoro generators are then differential
operators in the $a_k$'s. See ref.\cite{BBslevir}.

Concretely, the $SLE_{\kappa}$ equation $\dot{g}_t(z)=\frac{2}{g_t(z)-\xi_t}$
turns into a hierarchy of ordinary differential equations for the
coefficients of the expansion of $g_t(z)$ at infinity. Writing 
$g_t(z)\equiv z(1+\sum_{i \geq 2}a_i z^{-i})$ and $a_1 \equiv -\xi_t$
leads to 
$$\left\{\begin{array}{rcl}  \dot{a}_2 & = & 2 \\
\dot{a}_j & = & -\sum_{i=1}^{j-2} a_i\dot{a}_{j-i},\quad j \geq
3.\end{array} \right.$$ 

Define polynomials $p_j$ in the variables $a_i$ by
$$\left\{\begin{array}{rcl}  p_1 & = & 0 \\ p_2 & = & 1 \\
p_j & = & -\sum_{i=1}^{j-2} a_ip_{j-i},\quad j \geq 3. \end{array} \right.$$
Then by construction
$$\dot{a}_j=2p_j(a_1,\cdots), \quad j \geq 2.$$
If one assigns degree
$i$ to $a_i$, $p_j$ is homogeneous of degree $j-2$.  Using the fact
that $a_1(t)$ is a Brownian motion and Ito's formula, one derives a
general formula of Fokker-Planck type to compute the expectation of
any (polynomial, say) function $q(a_1(t),a_2(t),\cdots)$ :
$$\mathbf{E}[q(a_1(t),a_2(t),\cdots]=\left(e^{t\hat{A}}q(a_1,a_2,\cdots)
\right)_{0=a_1=a_2=\cdots}$$
where $A$ is the differential operator 
\debut
A=\frac{\kappa}{2}\frac{\partial^2}{\partial a_1 \, ^2}+2\sum_{j
  \geq 2}p_j\frac{\partial}{\partial a_j}.
\label{Apert}
\fin

In fact, $A$ is yet another avatar of the evolution
operator ${\cal A} = -2 L_{-2} +\frac{\kappa}{2}\, L_{-1}^2$
from eq.(\ref{slehamil}), this time acting on polynomials of the
$a_i$'s. To check this, remember that $w=f_t(z)$ describes the
integral curve starting from $z$ at $t=0$ for the time dependent
meromorphic vector field $\frac{2}{w}\partial_w
-\dot{\xi_t}\partial_w$. 
Then by definition $a_i=\oint w(z)z^{i-2}dz$ and the vector field
$-w^{n+1} \partial_w$ acts as
$$
\delta_n a_i=\oint \delta_n w(z)z^{i-2}dz=-\oint w(z)^{n+1}z^{i-2}dz.
$$
For $n \leq -1$ the right hand side vanishes for $i \leq 0$ which is a
consistency condition. One checks that $\delta_{-1} a_i=-\delta_{i,1}$ 
and $\delta_{-2} a_i=-p_i \quad i \geq 2$, so on 
polynomials in the variables $a_1,a_2,\cdots$, $L_{-1}$ acts as
$-\frac{\partial}{\partial a_1}$ and $L_{-2}$ as $-\sum_{j
  \geq 2}p_j\frac{\partial}{\partial a_j}.$

Using the representation (\ref{Apert}) for the stochastic evolution
operator $-2 L_{-2} +\frac{\kappa}{2}\, L_{-1}^2$, one may show that
polynomial martingales -- which are polynomials in the $a_i$ in the
kernel of $A$ --  are in one-to-one correspondence with states in $\CH_{1,2}$.
Namely, let us assign degree $j$ to $a_j$ and set ${\rm dim}_q
{\rm Ker}A=\sum_{n\geq 0}q^n D(n)$ with $D(n)$ the number of independent
homogeneous polynomials of degree $n$ in ${\rm Ker}A$. Then:
$$
{\rm dim}_q {\rm Ker} A = \frac{1-q^2}{\prod_{n\geq 1}(1-q^n)}
$$
This coincides with the graded character of $\CH_{1,2}$.
Furthermore, polynomials in $a_1,a_2,\cdots$ do not vanish when the
variables $a_i$ are replaced by their explicit expressions in terms of
the Brownian motion as obtained by solving Loewner's equation. 
This indicates that $G_t\ket{\omega}$ is a universal martingale 
-- meaning that it contains all martingales.
See ref.\cite{BBslevir} for more details.
 
\vskip 1.0 truecm

\section{Geometrical interpretation and 2D gravity.}
The aim of this section is to explain that the state $G_t\ket{\omega}$ 
involved in the definition of the above martingales possesses a simple
geometrical interpretation.

For each realization, the domain ${\bf H_t}={\bf H}\setminus K_t$ is
conformally equivalent to the upper half plane.  A boundary CFT
defined over it is coupled to the random $SLE_\kappa$ hulls.
Correlation functions of boundary operators $\Psi_{\delta_p}(x_p)$ and
of bulk operators $\Phi_{h_j}(z_j,\bar z_j)$ (with identical left and
right conformal dimensions $h_j=\bar h_j$ to keep formul\ae\ 
readable), are naturally defined by: 
\debut 
&& \vev{
  \prod_p\Psi_{\delta_p}(x_p) \cdot
  \prod_j \Phi_{h_j}(z_j,\bar z_j) }_{\om;\bf H_t} \label{corr1}\\
&& ~~~~~~~~~~~ \equiv \bra{0} \prod_p\Psi_{\delta_p}(x_p) \cdot
\prod_j\Phi_{h_j}(z_j,\bar z_j)\cdot G_t \ket{\omega} \non 
\fin 
where
the subscript ${\om;\bf H_t}$ refers to the domain of definition and
to the boundary conditions imposed to the conformal field theory, as
we shall discuss. The r.h.s. are products of intertwiner operators and
$\bra{0}$ is the conformally invariant dual vacuum with $\bra{0}L_n=0$
for $n\leq 1$.  Using eq.(\ref{gonpsi}) to implement the conformal
transformations represented by $G_t$, these correlators may also be
presented as \debut \bra{0}{ \prod_p[\mony_t'(x_p)]^{\delta_p}
  \Psi_{\delta_p}(\mony_t(x_p)) \cdot \prod_j |\mony_t'(z_j)|^{2h_j}
  \Phi_{h_j}(\mony_t(z_j),\bar \mony_t(\bar z_j))} \ket{\omega}
\label{corr2}
\fin
Although natural as a definition for a CFT over ${\bf H_t}$, 
the peculiarity of eq.(\ref{corr2}) is that the conformal
transformations $G_t$ do not act on the state $\ket{\omega}$
\footnote{This is actually necessary as the transformations
  $\mony_t(z)$ are singular at the origin.}.

\subsection{The hull boundary state.}
We first present an interpretation in terms of statistical models of the
occurrence of the state $\ket{\omega}$ with conformal weight
$h_{1,2}^\kappa$. We then give a meaning to the state
$G_t\ket{\omega}$ in terms of boundary CFT.

Let us recall a few basic elements concerning the microscopic
definition of the lattice statistical models that are believed to  
underlie $SLE_\kappa$ processes.
Consider for instance the $Q$-state Potts model, defined over some
lattice, whose partition functions is:
$$
Z= \sum_{\{s({\bf r})\}}\ \exp[\, J\sum_{{\bf r}\sqcup {\bf r'}}
\delta_{s({\bf r}),s({\bf r'})}\,]
$$
where the sum is over all spin configurations and the symbol
${\bf r}\sqcup {\bf r'}$ refers to neighbor sites ${\bf r}$ and
${\bf r'}$ on the lattice. The spin $s({\bf r})$ at
sites ${\bf r}$ takes $Q$ possible values from $1$ to $Q$. 
By expanding the exponential factor in the above expression
using $\exp[J(\delta_{s({\bf r}),s({\bf r'})}-1)]=(1-p) +
p\delta_{s({\bf r}),s({\bf r'})}$ with $p=1-e^{-J}$,
the partition function may be rewritten following Fortuin-Kastelyn
\cite{FK} as a sum over cluster configurations
$$
Z=e^{JL}\ \sum_C\, p^{\|C\|}\,(1-p)^{L-\|C\|}\, Q^{N_C}
$$
where $L$ is the number of links of the lattice, $N_C$  the number
of clusters in the configuration $C$ and 
$\|C\|$ the number of links inside the $N_C$ clusters.
In each of these so-called FK-clusters all spins take arbitrary but
identical values.

\begin{figure}[htbp]
  \begin{center}
    \includegraphics[width=0.9\textwidth]{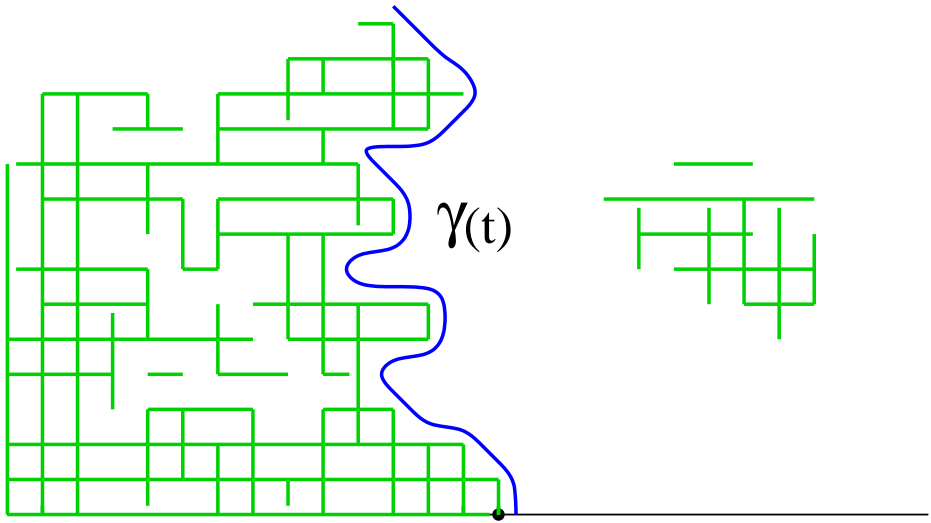}
      \caption{\em A FK-cluster configuration in the Potts models.
        The $SLE_\kappa$ trace $\gamma(t)$ is the boundary of the
        FK-cluster connected to the negative real axis.}
      \label{fig:slepotts}
  \end{center}
\end{figure}

Imagine now considering the $Q$-state Potts models on a lattice
covering the upper half plane with boundary conditions on the real
line such that all spins at the left of the origin are frozen to the
same identical value while spins on the right of the origin are free
with non assigned values. In each FK-cluster configuration there
exists a cluster growing from the negative half real axis into the
upper half plane whose boundary starts at the origin.  In the
continuum limit, this boundary curve is conjectured to be
statistically equivalent to a $SLE_\kappa$ trace
\cite{schramm0,schramm}. See Figure (\ref{fig:slepotts})\footnote{We
  thank A. Kupianen for his explanations concerning this point.}.

\bigskip

\bigskip

Being the boundary of a FK-cluster, the spin boundary conditions on
both sides of the $SLE_\kappa$ trace are not identical. Indeed, from
the above microscopic description we infer that in one side the 
spins are fixed whereas in the other side they are free.  This
change in boundary conditions corresponds to the insertion of a boundary
changing operator \cite{cardy} which is naturally identified with the
operator $\Psi_\om$ creating the state $\omega$:
$$
\Psi_\om(0)\ket{0} = \ket{\omega}
$$
with $\ket{0}$ the vacuum of the CFT defined over ${\bf H}$.
Note that creating a state at the origin in ${\bf H}$ corresponds
indeed to creating a state at the tip of the $SLE_\kappa$ trace
since $\gamma(t)=\mony_t^{-1}(0)$.
This explains the occurrence of the state $\ket{\omega}$ in
eq.(\ref{corr1},\ref{corr2}).

The relation between $Q$ and $\kappa$ can be found by matching
the known value of the dimension of the boundary changing operator for
the $Q$-state Potts model
with  $h_{1,2}^\kappa$. One gets:
$$ Q = 4\, \cos^2\big( \frac{4\pi}{\kappa}\Big), \quad 
\kappa \geq 4 $$
$Q=1$ for $\kappa=6$ as it should be for percolation,
and $Q=2$ with $\kappa=16/3$ for the Ising model.

\vskip 0.5 truecm

\begin{figure}[htbp]
  \begin{center}
    \includegraphics[width=0.9\textwidth]{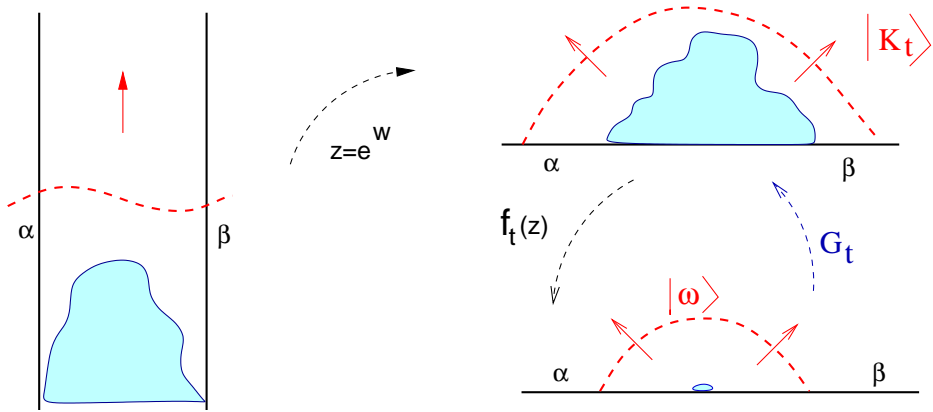}
      \caption{\em A representation of the boundary hull state and of
        the maps between different formulations of CFTs. The
        shaded domains represent the $SLE_\kappa$ hulls and the dashed 
        lines the `time' slices used to define CFT Hilbert spaces.}
      \label{fig:slehilbert}
  \end{center}
\end{figure}

To formulate the boundary CFT coupled to the $SLE_\kappa$ hulls,
imagine mapping back the domain ${\bf H_t}$ away from the hulls to a
strip of width $\pi$ such that half circles are mapped onto straight
lines. The map is $z=e^{\sigma+i\theta} \to w=\log z$ with
$\theta\in[0,\pi]$. There is a natural operator formalism for the CFT
defined on the strip, which is usually referred to as the open string
formalism, see eg. ref.\cite{Dif} chapters 6 and 11.  Boundary
conditions on the two sides $\theta=0$ and $\theta=\pi$ of the strip
are those inherited from the boundary conditions on the two sides of
the $SLE_\kappa$ traces that we just described. Constant `time'
slices, over which are defined conformal Hilbert spaces, are ${\rm
  Re}\,w\equiv\sigma={\rm const.}$ with the associated hamiltonian
operators generating evolutions in the $\sigma$ direction.  Under this
map, the $SLE_\kappa$ hulls appear as disturbances localized in the
far past region, with $\sigma$ small enough, thus generating states
belonging to the conformal Hilbert spaces. See Figure
(\ref{fig:slehilbert}).  Similarly, the domain ${\bf H_t}$ is mapped
onto the upper half plane using the by now familiar transformation
$z\to \mony_t(z)$ sending the tip of the $SLE_\kappa$ traces to the
origin. Under compositions of these two maps, the images of the
$\sigma={\rm const.}$ slices are curves topologically equivalent to
half circles around the origin. The `time' quantization valid in the
strip is replaced by the `radial' quantization with conformal Hilbert
spaces defined over these curves or over the constant radius half
circles.  The $SLE_\kappa$ hulls, which are then local perturbations
localized around the origin, generate states, which we denote
$\ket{K_t}$, in the radial quantization Hilbert spaces.  From
eq.(\ref{gonpsi}), we learn that $G_t$ intertwins the conformal field
theories defined over ${\bf H_t}$ and ${\bf H}$ so that
$G^{-1}_t\ket{K_t}$ is constant under $SLE_\kappa$ evolution.  This
leads to the identification 
\debut 
\ket{K_t}= G_t\, \ket{\omega}
\label{kate}
\fin
where we introduce the state $\ket{\omega}$ to keep track of the
boundary conditions we described above. 
Of course, up to the particular role played by the state
$\ket{\omega}$, this identification only reflects the fact that
conformal correlation functions on ${\bf H_t}$ are defined 
via the conformal map $\mony_t(z)$ represented by $G_t$.

This explains our identification (\ref{corr1}).

\bigskip\bigskip
  
\bigskip

\subsection{SLE and 2D gravity.}
We may interpret conformal field theories coupled to $SLE_\kappa$ hulls
as CFT's on random environments either by viewing them as 
CFT on random domains ${\bf H_t}$ but with a trivial metric, or by
viewing  them in the fixed domain ${\bf H}$, the upper half plane,
but with random metrics $|\mony'_t(z)|^2 dzd\bar z$ at points $\mony_t(z)$.
These points of view are similar to those adopted in studying 2D
gravity.

The conservation law (\ref{conserv}) then translates into a statement 
concerning the averaged conformal correlators
in these random environments. Namely:
$$
{\bf E}[\,\vev{ \prod_p\Psi_{\delta_p}(x_p) \cdot 
\prod_j \Phi_{h_j}(z_j,\bar z_j) }_{\om;\bf H_t}\, ]
= \bra{0}\, \prod_p\Psi_{\delta_p}(x_p) \cdot 
\prod_j \Phi_{h_j}(z_j,\bar z_j)\,\ket{\om}
$$

There is an intriguing, but interesting, link between the 
influence of the $SLE_\kappa$ hulls on conformal  scaling
properties and the KPZ formula for gravitationally
dressed dimensions \cite{kpz,ddk}.
As we are going to argue below, in average, scaling properties of
a boundary field $\Psi_\delta(x)$ of dimension $\delta$ close to the
hulls are dressed such that: 
\debut 
{\bf E}[\, \vev{\,\cdots
  \,\Psi_\delta(x)}_{\om;\bf H_t}\, ]\ \simeq _{x\to 0}\ 
{\rm const.}\ x^\Delta
\label{slekpz}
\fin
where $\cdots$ refers to the insertion of any operators away from the
hulls. The small $x$ limit mimics the approach of the field 
$\Psi_\delta(x)$ close to the hull since $\gamma(t)=\mony_t^{-1}(0)$.
 The dressed dimension $\Delta$ is determined by fusion rules and 
is solution of the quadratic equation
\debut
\Delta\Big({\Delta - \frac{\kappa-4}{\kappa} }\Big) =
\frac{4}{\kappa}\, \delta
\label{kpz00}
\fin
with solutions
\debut
\Delta_\pm(\delta) = \inv{2\kappa}\Big(\ \kappa-4 \pm 
\sqrt{(\kappa-4)^2+16\kappa \delta}\ \Big)
\label{kpzsol}
\fin
As a consequence, eq.(\ref{kpzsol}) also determines the conformal
weights of the intermediate states propagating in the correlation
functions we shall consider in the following Sections.

Eq.(\ref{slekpz}) may be derived as follows.
From eq.(\ref{conserv}) we have in average:
$$
{\bf E}[\, \vev{\,\cdots \,\Psi_\delta(x)}_{\om;\bf H_t}\, ]
=\bra{0}\, \cdots\,  \Psi_\delta(x)\,\ket{\om}
$$
For $x\to 0$, the leading contribution to the r.h.s. comes from the
highest weight vector, say $\ket{\alpha}$, created by $\Psi_\delta(x)$
acting on $\ket{\omega}$. Hence,
$$
\bra{0}\, \cdots\,  \Psi_\delta(x)\,\ket{\om} \simeq 
\bra{0}\, \cdots\,\ket{\alpha}\, \bra{\alpha}
\Psi_\delta(x)\,\ket{\om}
$$
As recalled in Appendix B, the null vector relation
$(-2L_{-2}+\frac{\kappa}{2}L_{-1}^2)\ket{\om}=0$ imposes constraints
on the possible states $\ket{\alpha}$ and on the possible scalings
of $\bra{\alpha}\Psi_\delta(x)\,\ket{\om}$. One has:
$$
\Big(\, 2\ell_{-2}^\delta(x) 
+\frac{\kappa}{2} \ell_{-1}^{\delta}(x)^2\,\Big)
\bra{\alpha}\Psi_\delta(x)\,\ket{\om}=0
$$
This gives $\bra{\alpha}\Psi_\delta(x)\,\ket{\om}={\rm const.}\,
x^\Delta$ with $\Delta$ solution of the eq.(\ref{kpz00}).

Eq.(\ref{kpz00}) is the famous KPZ relation linking scaling
dimension in flat space with that dressed by 2D gravity
\cite{kpz,ddk}, at least for $\kappa\leq 4$.
The KPZ relation is usually written as
$$
\Delta(\Delta - \gamma_{\rm str.})=(1-\gamma_{\rm str.})\,\delta
$$
where $\gamma_{\rm str.}$ takes one of the two possible values
$$\gamma_{\rm str.}^\pm=\Big(\,c-1\pm\sqrt{(1-c)(25-c)}\,\Big)/12.$$
The choice $\gamma_{\rm str.}^-$ is usually picked on based on
semi-classical arguments valid for $c\to-\infty$. Being a function of the
central charge, $\gamma_{\rm str.}^\pm$ is invariant by duality,
$\gamma_{\rm str.}^\pm(\kappa)=\gamma_{\rm str.}^\pm(\hat \kappa)$
for $\hat \kappa=16/\kappa$. One has:
$$
\gamma_{\rm str.}^-(\kappa)=\frac{\kappa-4}{\kappa},\quad
\gamma_{\rm str.}^+(\kappa)=\frac{4-\kappa}{4};\quad 
\kappa \leq 4
$$
Hence, eq.(\ref{kpz00}) matches with the KPZ equation with
$\gamma_{\rm str.}=\gamma_{\rm str.}^-$ for $\kappa\leq 4$.  To 
reconciliate the two equations for $\kappa\geq 4$ one may try to invoke
that the other branch for $\gamma_{\rm str.}$ has to be used above the
selfdual point.

The fact that the scaling relation (\ref{slekpz}) reproduces that
predicted by the KPZ relation suggests 
that, at least for $\kappa\leq 4$,
the $SLE_\kappa$ average samples a part of the 2D gravity phase
space large enough to test -- and to exhibit -- scaling behaviors
in presence of gravity.
It would be interesting to turn the above observation in more
complete statements.

\vskip 1.0 truecm

\section{SLE crossing probabilities from CFT.}
In this section we show on a few examples how to use the martingale
properties (\ref{martin},\ref{stop}) to compute crossing
probabilities first computed refs.\cite{LSW,schramm0,another}.

The approach consists in choosing, in an appropriate way
depending on the problem, the vector $\bra{v}$ on which
eq.(\ref{stop}) is projected such that the expectation 
$${\bf E}[\bra{v}\, G_\tau\ket{\omega}]=\langle{v}\ket{\omega}$$ 
may be computed in a simple way in
terms of the crossing probabilities. 
In other words, given an event ${\cal E}$ associated to a stopping
time $\tau$, we shall identify a vector $\bra{v_{\cal E}}$ such
that $$ \bra{v_{\cal E}}\, G_\tau\ket{\omega}= {\bf 1}_{\cal E}$$
This leads to linear systems for  
these probabilities whose coefficients are correlation functions of a
conformal field theory defined over the upper half plane.
Although our approach and that of refs.\cite{LSW,schramm0,another} are
of course linked, they are in a way reversed one from the other.
Indeed, the latter evaluate these crossing
probabilities using the differential equations they satisfy -- because 
they are associated to martingales, while we compute them by
identifying them with CFT correlation functions -- because they are
associated to martingales -- and as such they satisfy the differential 
equations. For the events ${\cal E}$ we shall consider below, the
vectors $\bra{v_{\cal E}}$ are constructed using conformal fields.
Since we use operator product expansion properties \cite{bpz} of
conformal fields to show that the vectors $\bra{v_{\cal E}}$ satisfy
the appropriate requirements, our apporach is more algebraic than that 
of refs.\cite{LSW,schramm0,another} but less rigorous.

\subsection{Generalized Cardy's formula.}
The most famous crossing probability is that of Cardy \cite{cardy2}
which gives the probability that there exists a percolating cluster
in critical percolation connecting to opposite sides of a rectangle. 

Cardy's formula for critical percolation applies to $SLE_6$.  It may
be extended \cite{LSW} to a formula valid in $SLE_\kappa$ for any
$\kappa>4$. The problem is then formulated as follows.  Let
$-\infty<a<0<b<\infty$ and define the stopping times $\tau_a$ and
$\tau_b$ as the first times at which the $SLE_\kappa$ trace
$\gamma(t)$ touches the interval $(-\infty,a]$ and $[b,+\infty)$
respectively:
\debut
\tau_a &=& {\rm \inf}\{ t>0;\ \gamma(t)\in (-\infty,a]\}\non\\
\tau_b &=& {\rm \inf}\{ t>0;\ \gamma(t)\in [b,+\infty)\}\non
\fin
By definition, Cardy's crossing probability is the probability that
the trace hits first the interval $(-\infty,a]$, that is:
$$
{\bf P}[\,\tau_a<\tau_b\,]
$$

Let $F_t^{(1)}(a,b)$ be the following correlation function
$$ 
F_t^{(1)}(a,b)\equiv \bra{\omega}
\Psi_{\delta=0}(a)\Psi_{\delta=0}(b)G_t\ket{\omega} 
$$
with $\Psi_{\delta=0}(x)$ a boundary conformal field of scaling
dimension zero and $\bra{\omega}$ the dual of the highest
weight vector $\ket{\omega}$. There exist actually two linearly 
independent correlators, one of them being constant, but we shall not
specify yet which non constant correlation function we pick.  Recall
that $G_{t=0}=1$.  By dimensional analysis -- or because of the
commutation relations with $L_0$ -- $F_{t=0}^{(1)}(a,b)$ is only a
function of the dimensionless ratio $a/(a-b)$:
$$
F_{t=0}^{(1)}(a,b) = \tilde F_{\delta=0}(u)\ ,\quad u\equiv\frac{a}{a-b}
$$
The cross ratio $u$, $0<u<1$, has a simple interpretation: it is the
image of the origin, the starting point of the $SLE_\kappa$ trace,
 by the homographic transformation fixing the
infinity and mapping the point $a$ to $0$ and $b$ to $1$.

We apply the martingale property (\ref{stop}) for the stopping
time $\tau={\rm min}(\tau_a,\tau_b)$ so that: 
\debut 
{\bf E}[\, F_\tau^{(1)}(a,b)\,] = F_{t=0}^{(1)}(a,b) =
\tilde F_{\delta=0}(u)
\label{basicardy} 
\fin 
Similarly as in eq.(\ref{bigbig}) we compute $F_\tau^{(1)}(a,b)$ 
by using the intertwining relation (\ref{gonpsi}) 
to move $G_\tau$ to the left of the correlators. This gives:
$$
 F_\tau^{(1)}(a,b)= F_{t=0}^{(1)}( \mony_\tau(a),\mony_\tau(b) )
= \tilde F_{\delta=0}(u_\tau)\ ,\quad
u_\tau=\frac{\mony_\tau(a)}{\mony_\tau(a)-\mony_\tau(b)}
$$
The important observation is that $u_\tau$ takes two simple 
non-random values depending whether $\tau$ equals $\tau_a$ or $\tau_b$:
\debut
\tau=\tau_a \Leftrightarrow \tau_a<\tau_b&:&\quad u_\tau=0 \label{utau}\\
\tau=\tau_b \Leftrightarrow \tau_a>\tau_b&:&\quad u_\tau=1 \non
\fin
Indeed, $\tau=\tau_a$ means that the trace $\gamma(t)$ hits first the interval
$(-\infty,a]$. As illustrated in the example of Appendix A, 
at the instance at which the trace gets back to the real axis, all
points which have been surrounded by it are mapped to the origin by $f_t(z)$.
Therefore, if $\tau=\tau_a$ then $\mony_\tau(a)=0$ while $\mony_\tau(b)$
remains finite and thus $u_\tau=0$. The case $\tau=\tau_b$ is
analysed similarly. Thus
$$
F_\tau^{(1)}(a,b)= {\bf 1}_{\{\tau_a<\tau_b\}}\, \tilde F_{\delta=0}(0)
+{\bf 1}_{\{\tau_a>\tau_b\}}\, \tilde F_{\delta=0}(1)
$$
and we can compute ${\bf E}[F_\tau^{(1)}(a,b)]$ in terms
of ${\bf P}(\tau_a<\tau_b)$:
$$
{\bf E}[F_\tau^{(1)}(a,b)]= {\bf P}[\,\tau_a<\tau_b\,]\, \tilde F_{\delta=0}(0)
+ (1-{\bf P}[\,\tau_a<\tau_b]\,)\, \tilde F_{\delta=0}(1)
$$
where we used that ${\bf P}[\,\tau_a>\tau_b\,]=1-{\bf P}[\,\tau_a<\tau_b\,]$.
Together with the basic martingale equation (\ref{basicardy}) we get
Cardy's formula: 
\debut
{\bf P}[\,\tau_a<\tau_b\,] = 
\frac{\tilde F_{\delta=0}(u)-\tilde F_{\delta=0}(1)}{
\tilde F_{\delta=0}(0)-\tilde F_{\delta=0}(1)}
\label{cardyform}
\fin
The correlation function $\tilde F_{\delta=0}(u)$ is shown below to be a
hypergeometric function.
\vskip 0.5 truecm

Cardy's formula can be further generalized \cite{LSW} by considering 
for instance the expectation:
$$
{\bf E}[\, {\bf 1}_{\{\tau_a<\tau_b\}}\, 
\Big(\frac{\mony_\tau'(b)}{\mony_\tau(b)-\mony_\tau(a)}\Big)^\delta\,]
$$
This may be computed as above but considering correlation functions with 
the insertions of two boundary operators, one of dimension zero and the 
other one of dimension $\delta$. Namely
\footnote{To be precise, the state $\bra{\omega}$ in
  eq.(\ref{Fdeux}) belongs, for generic $\delta$, to the dual space of
  the contravariant representation $\tilde {\cal V}_{h_{1,2}^\kappa}$,
  cf. Appendix B. We however still denote it by $\bra{\omega}$ to
  avoid cumbersome notations.}:
\debut 
F_t^{(2)}(a,b)\equiv\bra{\omega} 
\Psi_{\delta=0}(a)\Psi_{\delta}(b)G_t\ket{\omega}
\label{Fdeux}
\fin
Again there exist two independent such correlators, corresponding to two different
conformal blocks \cite{bpz}, and we shall specify in a while
which one we choose. See Appendix B for further details.
Since $G_{t=0}=1$, dimensional analysis tells that:
$$
F_{t=0}^{(2)}(a,b)= \inv{(b-a)^\delta}\, \tilde F_\delta(u),\quad
u=\frac{a}{a-b}
$$
As above, we start from the martingale relation (\ref{stop}):
$$
{\bf E}[\,F_\tau^{(2)}(a,b) \,] = F_{t=0}^{(2)}(a,b)
$$
with $\tau={\rm min}(\tau_a,\tau_b)$.
Moving $G_\tau$ to the left using the intertwining relation
(\ref{gonpsi}) gives:
$$
F_\tau^{(2)}(a,b)=
\Big(\frac{\mony_\tau'(b)}{\mony_\tau(b)-\mony_\tau(a)}\Big)^\delta\
\tilde F_\delta(u_\tau),\quad 
u_\tau=\frac{\mony_\tau(a)}{\mony_\tau(a)-\mony_\tau(b)}
$$
The difference with the previous computation for Cardy's formula is
the occurrence of the Jacobian $\mony_t'(b)$ since $\delta$ is non vanishing.
The argument (\ref{utau}) concerning the possible values of $u_\tau$
still applies. So, choosing the correlation function $ \tilde F_\delta(u)$
which vanishes at $u=1$ selects the case $\tau_a<\tau_b$ as the only
contribution to the expectation ${\bf E} [\,F_\tau^{(2)}(a,b) \,]$.
Hence:
\debut
\tilde F_\delta(0)\ {\bf E}[\, {\bf 1}_{\{\tau_a<\tau_b\}}\, 
\Big(\frac{\mony_\tau'(b)}{\mony_\tau(b)-\mony_\tau(a)}\Big)^\delta\,]
= \inv{(b-a)^\delta}\, \tilde F_\delta(u)
\label{deuxpoints}
\fin
with $\tilde F_\delta(u=1)=0$. This agrees with ref.\cite{LSW}.
Choosing different boundary conditions for $\tilde F_\delta(u)$ would
give different weights to the events $\{\tau_a<\tau_b\}$ and
$\{\tau_a>\tau_b\}$. 

As rederived in Appendix B, the correlation function
$\bra{\omega} \Psi_{\delta=0}(a)\Psi_{\delta}(b)\ket{\omega}$
satisfies a differential equation as a consequence of the existence of
the null vector $\ket{n_{1,2}}$ at level two in the Verma module generated
over $\ket{\om}$. For the function $\tilde F_\delta(u)$ this differential
equation translates into:
\debut
\Big[\, -4\delta + 4(u-1)(2u-1)\partial_u + \kappa u(u-1)^2\partial_u^2
\,\Big]\, \tilde F_\delta(u) =0 \non
\fin
The two solutions correspond to the two possible conformal blocks, and
as discussed above we select one of them by demanding $\tilde
F_\delta(1) =0$. As it is well known,
the solutions may be written in terms of hypergeometric functions.

Notice that the states propagating in the intermediate channel of the
correlation $\bra{\omega}
\Psi_{\delta=0}(a)\Psi_{\delta}(b)\ket{\omega}$ are those created by
$\Psi_\delta$ acting on $\ket{\omega}$ with conformal weights
$\Delta_\pm(\delta)+\delta+h_{1,2}^\kappa$, eq.(\ref{kpzsol}), as
explained in Appendix B.  As $b\to 0$, or equivalently $u\to 1$, the
leading contributions are governed by these intermediate states so
that:
$$
\bra{\omega} \Psi_{\delta=0}(a)\Psi_{\delta}(b)\ket{\omega}
\simeq_{b\to0} b^{\Delta_\pm(\delta)}\, C_{\delta\om}^{h_\pm(\delta)}\,
\bra{\omega} \Psi_{\delta=0}(a)\ket{h_\pm(\delta)}+\cdots
$$
with $h_\pm(\delta)=\delta+h^\kappa_{1,2}+\Delta_\pm(\delta)$
and $C_{\delta\om}^{h_\pm(\delta)}$ the structure constants.
As recalled in Appendix B, the two possibilities correspond to the two
possible choices of intertwiners acting on $\CH_{1,2}$.
Since $\Delta_+(\delta)>0$ while $\Delta_-(\delta)<0$  for $\delta>0$, 
imposing $\tilde F_\delta(1)=0$ selected the conformal block creating
the states $\ket{h_+(\delta)}$, 
ie $\Psi_\delta(x):\CH_{1,2}\to \CV_{h_+(\delta)}$, for $\delta>0$.

\vskip 0.5 truecm

\subsection{Boundary excursion probability.}

As another application of the conformal machinery to a probabilistic
stopping time problem, consider the following question, already
answered by standard methods in ref.\cite{schramm}. 
If $\gamma(t)$ is an $SLE_\kappa$ path, $\kappa
\in ]4,8[$, and $u$ a point on the positive real axis, let $x=\inf
\{[u,+\infty[ \cap \gamma[0,+\infty[\}$ be the first point $\geq u$
touched by the $SLE_\kappa$ path. What is the distribution of $x$ ? This
question is similar to the one answered by Cardy's formula. Note
that in fact, $x$ is the position of the $SLE_\kappa$ trace at time
$\tau_u$, the last time for which the Loewner map is well-defined at $z=u$.

To prepare for the answer to this question, we study the two point
correlator 
$$
\bra{\omega} \Psi_{\delta=0}(v)\Psi_{\delta=0}(u)G_t\ket{\omega}
$$
for $0<u < v < +\infty$. 

First, suppose that $t=0$. 
If $u$ comes close to $0$, we can expand this function by computing
the operator product expansion of
$\Psi_{\delta=0}(u)\ket{\omega}$. As already discussed at
length, this can involve at most two conformal
families. The conformal family of $\ket{\omega}$ could be one of
these but we shall remove it by demanding that
it does not appear in the operator product expansion. This fixes the
boundary conditions we shall impose on the correlation functions. Then
$\Psi_{\delta=0}(u)\ket{\omega} \sim u ^{\frac{\kappa -4}{\kappa}}
\ket{\Psi_{\delta=\frac{\kappa -2}{2\kappa}}}$. This goes to $0$ iff
$\kappa > 4$

If the points $u$ and $v$ come close together, the operator product expansion 
$\Psi_{\delta=0}(v)\Psi_{\delta=0}(u)$ is more involved. General rules of
conformal field theory ensure that the identity operator contributes,
but apart from that, there is no a priori restrictions on the
conformal families $\Psi_{\delta}$ that may appear. However, only
those for which $\bra{\omega} \Psi_{\delta}\ket{\omega}\neq 0$
remain, and this restricts to two conformal families, the identity and
$\Psi_{\delta=h_{1,3}^\kappa}$. See Appendix B. 
Thus, when $u$ and $v$ come close together,
the dominant contribution to $\bra{\omega}
\Psi_{\delta=0}(v)\Psi_{\delta=0}(u)\ket{\omega}$ is either
$1$ or
$(v-u)^{\frac{8-\kappa}{\kappa}}\bra{\omega}
\Psi_{\delta=h_{1,3}^\kappa}\ket{\omega}$ depending on
whether $\kappa<8$ or $\kappa>8$.

Hence, if $\kappa \in ]4,8[$, the correlation function $
\bra{\omega} \Psi_{\delta=0}(v)\Psi_{\delta=0}(u)\ket{\omega}
$ vanishes at $u=0$ and takes value $1$ at $u=v$. 

For nonzero $t$, we write  
\begin{eqnarray*}
\bra{\omega}
\Psi_{\delta=0}(v)\Psi_{\delta=0}(u)G_t\ket{\omega}
& = & \bra{\omega} \Psi_{\delta=0}(f_t(v)) 
\Psi_{\delta=0}(f_t(u)) \ket{\omega} \\
& = & \bra{\omega} \Psi_{\delta=0}(1) 
\Psi_{\delta=0}(f_t(u)/f_t(v)) \ket{\omega}.
\end{eqnarray*}
The last equality follows by dimensional analysis.  Now if $x$, the
position of the $SLE_\kappa$ trace at $t=\tau_u$, satisfies $u < x < v$,
$f_{\tau_u}(v)$ remains away from the origin but $f_{\tau_u}(u)=0$ and
the correlation function vanishes. On the other hand, if $v \leq x$,
it is a general property of hulls that $\lim _{t \nearrow
\tau_u}f_t(u)/f_t(v)=1$ and the correlation function is unity.

To summarize $\bra{\omega} \Psi_{\delta=0}(v)
\Psi_{\delta=0}(u) G_{\tau _u}\ket{\omega}=\mathbf{1}_{\{x \geq v\}}$.
From the martingale property (\ref{stop}) we infer that the
probability distribution function of $x$ is 
\debut
\mathbf{ E}[\mathbf{1}_{\{x \geq v\}}]=\bra{\omega}
\Psi_{\delta=0}(v)\Psi_{\delta=0}(u)\ket{\omega}.
\label{excur0}
\fin

As recalled in Appendix B, the fact that 
$$\bra{\omega}
\Psi_{\delta=0}(v)\Psi_{\delta=0}(u)(-2L_{-2}+
\frac{\kappa}{2}L_{-1}^2)\ket{\omega}=0$$ 
translates into the differential equation 
$$  \left(\frac{2}{u}\partial_u+\frac{2}{v}\partial_v +
\frac{\kappa}{2}(\partial_u+ \partial_v)^2   \right)\bra{\omega}
\Psi_{\delta=0}(v)\Psi_{\delta=0}(u)\ket{\omega}=0.
$$
Since the correlation function depends only on $s=u/v$,  we derive that
$$\left(\frac{d^2}{ds^2}+\left(\frac{4}{\kappa s} +
\frac{2(4-\kappa)}{\kappa(1-s)}\right)\frac{d}{ds}
\right)\bra{\omega} \Psi_{\delta=0}(1)\Psi_{\delta=0}(s)
\ket{\omega}=0.$$
The differential operator annihilates the
constants, a remnant of the fact that the identity operator has weight
0. With the normalization choosen for $\Psi_{\delta=0}$, the
relevant solution vanishes at the origin. The integration is
straightforward.

Finally, 
$${\bf P}[x \geq u/s]\equiv \mathbf{E}[\mathbf{1}_{\{x \geq u/s\}}] =
\frac{s^{\frac{\kappa -4}{\kappa}}\Gamma\left(\frac{4}{\kappa}
\right)}{\Gamma\left(\frac{\kappa-4}{\kappa}\right)\Gamma
\left(\frac{8-\kappa} {\kappa}\right)}\int_0^1 d\sigma \sigma
^{-\frac{4}{\kappa}} (1-s\sigma)^{2\frac{4-\kappa}{\kappa}}$$

This example is instructive, because it shows in a fairly simple case
that the thresholds $\kappa=4,8$ for topological properties of $SLE_\kappa$
appear in the CFT framework as thresholds at which divergences emerge
in operator product expansions.

\subsection{Domain wall probability.}
We now present another percolation formula  involving
bulk operators. For the underlying statistical model defined 
on the upper half plane as in Section 4, this
formula gives the probability for the position of the domain wall
limiting the FK-cluster connected to the negative real axis
relative to a given point in ${\bf H}$.
It was first proved in \cite{another} using
the $SLE_\kappa$ formulation and it was not yet derived within conformal
field theory.

Let $\kappa<8$ and $z$ be a point in ${\bf H}$. One looks for the
probability that the $SLE_\kappa$ trace goes to the left or to the
right of that point. Following ref.\cite{another}, it is convenient to 
reformulate this problem as a statement concerning the behavior of the
$SLE_\kappa$ trace at the time $\tau_z$ at which the point $z$ is
swallowed by the hull:
$$
\tau_z = {\rm inf}\{ t>0;\ f_t(z)=0\}
$$
At time $t\to \tau_z^-$, the point $z$ is mapped to the origin by 
$f_t$. Furthermore, with $f_t(z)=x_t(z)+iy_t(z)$,
almost surely the ratio $x_t(z)/y_t(z)$ goes to
$\pm\infty$, as illustrated in Appendix A.  The sign
$\pm\infty$ depends whether the point is surrounded by the trace
$\gamma(t)$ from the right or from the left  -- by convention, we say
that the trace goes to the right if it loops around $z$ positively. Hence, if 
$\theta_z\equiv \lim_{t\to \tau_z^-} x_t(z)/y_t(z)$ then \cite{another}:
$$
{\bf P}[ \gamma(t)\ {\rm to~the~ left[right]~of}\ z]
={\bf P}[\theta_z=-\infty[+\infty]\,]
$$
with ${\bf P}[\theta_z=+\infty]+{\bf P}[\theta_z=-\infty]=1$.

We shall derive these probabilities using the martingale equation
(\ref{stop}) with the correlation function
\debut
C^{(1)}_t(z,\bar z)\equiv 
\bra{\omega} \Phi_{h=0}(z,\bar z)\,G_t \ket{\omega}
\label{Cun}
\fin
with $\Phi_{h=0}(z,\bar z)$ a bulk operator of scaling dimensions zero. 
As for Cardy's formula, we choose arbitrarily any non-constant such
correlators.
Since $G_{t=0}=1$, dimensional analysis implies that $C^{(1)}_t(z,\bar
z)$ only depends on a dimensionless ratio:
$$
C^{(1)}_{t=0}(z,\bar z) = \tilde C_{h=0}(x/y),\quad z=x+iy
$$
The martingale equation (\ref{stop}) applied to $C^{(1)}_t(z,\bar z)$
at $t\to\tau_z^-$ yields:
\debut
{\bf E}[\, C^{(1)}_{\tau_z}(z,\bar z)\,] = \tilde C_{h=0}(x/y).
\label{bulkmartin}
\fin
As above, $G_t$ may be moved to the left by using the intertwining
relation (\ref{gonpsi}) so that:
$$
C^{(1)}_{\tau_z}(z,\bar z)=
\lim_{t\to\tau_z}C^{(1)}_{t=0}(f_t(z),\bar f_t(\bar z))
=\tilde C_{h=0}(\theta_z)
$$
Since $\theta_z$ takes the two simple non-random values $\theta_z=\pm\infty$, 
the expectation ${\bf E}[\,C^{(1)}_{\tau_z}(z,\bar
z)\,]$ is computable:
$$
{\bf E}[\, C^{(1)}_{\tau_z}(z,\bar z)\,]= 
{\bf P}[\theta_z=+\infty]\, \tilde C_{h=0}(+\infty)+
{\bf P}[\theta_z=-\infty]\,\tilde C_{h=0}(-\infty)
$$
With the martingale equation (\ref{bulkmartin}), it implies:
\debut
{\bf P}[\theta_z=+\infty]=\frac{\tilde C_{h=0}(x/y)-\tilde
  C_{h=0}(-\infty)}{\tilde C_{h=0}(+\infty)-\tilde C_{h=0}(-\infty)}
\label{leftright}
\fin 
As we recall below, the function $\tilde C_{h=0}(x/y)$ satisfies
a second order differential equation and it is expressible as a
hypergeometric function. Eq.(\ref{leftright}) agrees with that of
ref.\cite{another}. 

\vskip .5 truecm

As with Cardy's formula, the above crossing probability admits a
simple generalization. Let us define
\debut
C^{(2)}_t(z,\bar z)\equiv \bra{\omega} \Phi_h(z,\bar z) G_t
\ket{\omega} \label{bulkcorr}
\fin
with $\Phi_h(z,\bar z)$ a bulk operator of scaling dimension $h$.
Scaling analysis implies that at initial time
$$
C^{(2)}_{t=0}(z,\bar z)=  y^{-2h\,} \tilde C_h(x/y),\quad z=x+iy
$$
As we are now familiar with, the martingale equation (\ref{stop}) for
$t\to\tau_z^-$ gives: 
\debut
{\bf E}[\, \Big(\frac{|\mony_{\tau_z}'(z)|}{{\rm
    Im}\mony_{\tau_z}(z)}\Big)^{2h}\, \tilde C_h(\theta_z)\,]
= y^{-2h}\,\tilde C_h(x/y)
\label{bigexpect}
\fin
with as before $\theta_z=\lim_{t\to\tau_z^-}x_t(z)/y_t(z)$.
Since $\theta_z$ takes only two values $\pm\infty$ depending
whether the trace loops positively or negatively around $z$,
eq.(\ref{bigexpect}) becomes:
\debut
&&{\bf E}[\, \Big(\frac{|\mony_{\tau_z}'(z)|}{{\rm
    Im}\mony_{\tau_z}(z)}\Big)^{2h}\, 
\Big({\bf 1}_{\{\theta_z=+\infty\}}\tilde C_h(+\infty)
+{\bf 1}_{\{\theta_z=-\infty\}}\tilde C_h(-\infty)\Big)\,] \non\\
&& ~~~~~~~~~~~~~~~~~~~~~~~~~~~~ = y^{-2h}\,\tilde C_h(x/y)
\label{unpoint}
\fin
There exist two conformal blocks for the correlation function
(\ref{bulkcorr}) and therefore two possible independent choices for
$C^{(2)}_t(z,\bar z)$. As a basis, one may select the even and odd
correlations $\tilde C_h(x/y)$.
Each choice leads to different
informations on the expectation (\ref{bigexpect}). 
The simplest is provided by choosing the even conformal block
so that the $\theta_z$ dependence in eq.(\ref{bigexpect}) factorizes.
This yields:
\debut
{\bf E}[\, \Big(\frac{|\mony_{\tau_z}'(z)|}{{\rm
    Im}\mony_{\tau_z}(z)}\Big)^{2h}\,]\, \tilde C_h^{[even]}(\infty)
= y^{-2h}\,\tilde C_h^{[even]}(x/y)
\fin

As recalled in Appendix B, $\tilde C_h(x/y)$ satisfies a
second order differential equation due to the existence of the null
vector $\ket{n_{1,2}}$ at level two. Its form is
$$
[\, 16h +8(1+\eta^2)\eta\partial_\eta+
\kappa(1+\eta^2)^2\partial^2_\eta\,]\tilde C_h(\eta)=0
$$
with $\eta=x/y$.  As usual its solutions are hypergeometric
functions.  To conclude for the expectation (\ref{bigexpect}) one has
to evaluate $\tilde C_h(\pm\infty)$, and to check whether it is
finite, zero or infinite. As discussed in ref.\cite{schramm}, this
depends on the values $h$ and $\kappa$. 

The different cases are
distinguished according to the conformal weights of the intermediate
states propagating in the correlations (\ref{bulkcorr}).
They may also be understood in terms of fusion rules.  The
limit $x/y\to\infty$ may be thought as the limit in which the bulk
operator $\Phi_h(z,\bar z)$ gets close to the real axis. By operator
product expansion, it then decomposes on boundary operators
$\Psi_\delta(x)$. Since we are taking matrix elements between
$\ket{\om}$ and $\ket{h_{1,2}^\kappa}$, only the operators with
 $\delta=0$ or $\delta=h_{1,3}^\kappa$ give non-vanishing
contributions, see Appendix B. Thus:
$$
y^{2h}\, \Phi_h(z,\bar z) \simeq_{y\to 0}
\mathbf{B}_h^{0}\, [ \Psi_{\delta=0}(x) +\cdots]+
y^{h_{1,3}^\kappa}\,\mathbf{B}_h^{h_{1,3}^\kappa}\, 
[\Psi_{\delta=h_{1,3}^\kappa}(x) +\cdots] +\cdots
$$
The different cases depend on the finiteness of the two bulk-boundary
coupling constants $\mathbf{B}_h^0$ and
$\mathbf{B}_h^{h_{1,3}^\kappa}$ and on the sign of
$h_{1,3}^\kappa=\frac{8-\kappa}{\kappa}$ -- that is whether $\kappa>8$
or $\kappa<8$.

\section{Exterior/interior hull relations.}

As the examples we have treated in detail show, the behavior of
conformal correlators when points are swallowed by the $SLE_\kappa$
hull is crucial for the probabilistic interpretation. In this section
we gather a few general remarks on this behavior.

Suppose that a domain $D$ is swallowed by the hull at time
$t=t_c$. Then it is known that
\begin{description}
\item[i)] in the unbounded component, the Loewner map $g_t$ has a limit. This
limit maps the unbounded component
conformally onto ${\bf H}$,
\item[ii)] in the bounded component $D$, the Loewner map has a limit which
is a constant map, so that the image of $D$ collapses to the point
$\xi_{t_c}$ on the real axis.
\item[iii)] for $z,z' \in  D$, 
$\lim _{t \nearrow t_c}\frac{g_{t}(z)-\xi _t}{g_{t}(z')-\xi _t}=1.$
\end{description}

Let us elaborate a little bit on property $iii)$. We concentrate on
the case when a simple path $\gamma_t$ starts from the origin at $t=0$
and touches the real axis at $t=t_c$ \footnote{In more complicated
  cases, the path has a self intersection at $t_c$ and/or has already
  made a (finite) number of self intersections before $t_c$, but
  choosing any $t_o$ such that $\gamma_t$ is simple for $t \in
  [t_o,t_c]$, the study of $g_t \circ g_{t_o}^{-1}$ instead of $g_t$
  reduces the problem to the situation that we treat in some detail.}.
Then the boundary of $D$ contains an open interval $I$ of the real
axis. Let $x$ be a point in this interval. By appropriate fractional
linear transformations on $z$ and $g_t(z)$, one can exchange the role
of the bounded and the unbounded component and then use property $i)$
\footnote{For instance, on may choose any $\tilde{x}$ such that
  $x\tilde{x}<0$ and set $\tilde{z}=\frac{\tilde{x}z}{z-x}$. The
  fractional linear transformation for $g_t(z)$ is then fixed by the
  normalization condition at infinity.}. This leads to
\begin{description}
\item[iv)] For $z \in  D$ and $x \in I$,
$\lim_{t \nearrow t_c} \frac{g_t(z)-g_t(x)}{g_t'(x)}$
exists and maps $D$ conformally onto ${\bf H}$. 
\end{description}

Put together, properties $ii)$, $iii)$ and $iv)$ imply that there
exist two functions of $t$, $\varepsilon^{(0)}_t=o(1)$ and
$\varepsilon^{(1)}_t=o(\varepsilon^{(0)}_t)$ such that 
$$\lim_{t \nearrow t_c} \frac{g_t(z)-\xi
  _t-\varepsilon^{(0)}_t}{\varepsilon^{(1)}_t}\equiv \check{g}_{t_c}(z)$$
exists and maps $D$ conformally onto ${\bf H}$.

For the semicircle example treated in detail in the appendix, one can
take $\varepsilon^{(0)}_t=-\sqrt{2(t_c-t)}$,
$\varepsilon^{(1)}_t=2(t_c-t)$ and
$\check{g}_{t_c}(z)=\frac{z}{(r+z)^2}$. 
In fact heuristic arguments
suggest that $\varepsilon^{(0)}_t \sim \pm \sqrt{2(t_c-t)}$ 
and $\varepsilon^{(1)}_t \sim 2(t_c-t)$ if the
curve $\gamma_t$ is smooth enough~\footnote{The
sign is $+$ if the boundary of $D$ is oriented clockwise by the path
and $-$ otherwise.}. 

This smoothness assumption is not
natural in the context of $SLE_\kappa$ so we work with
$\varepsilon^{(0)}_t$ and $\varepsilon^{(1)}_t$. 

Suppose there is a collection of fields in $\mathbf{H}$, 
with the convention that the ``$i(n)$'' arguments are inside $D$ or
$I$, and the ``$o(ut)$'' arguments sit outside.
Inserting resolutions of the identity and using covariance under
similarities as detailed in the Appendix C, we find that the
dominant contribution to the correlator
$$ \bra{0} 
\prod_{p_{o}}\Psi_{\delta_{p_{o}}}(x_{p_{o}})
\prod_{j_{o}}\Phi_{h_{j_{o}}}(z_{j_{o}},\bar z_{j_{o}})\cdot
\prod_{p_{i}}\Psi_{\delta_{p_{i}}}(x_{p_{i}}) 
\prod_{j_{i}}\Phi_{h_{j_{i}}}(z_{j_{i}},\bar z_{j_{i}})\cdot
G_t \ket{\alpha}$$
for $t$ close to $t_c$ is given by 
\begin{eqnarray}
\label{eq:grosse}
\sum_{\beta ,\eta}\mathbf{C}^{\beta }_{\alpha \; \eta}\,
(\varepsilon^{(0)}_t)^{\delta_{\beta}-\delta_{\alpha}-
  \delta_{\eta}}(\varepsilon^{(1)}_t)^{\delta_{\eta}}\,
\bra{0} \prod_{p_{o}}\Psi_{\delta_{p_{o}}}(x_{p_{o}}) 
\prod_{j_{o}}\Phi_{h_{j_{o}}}(z_{j_{o}},\bar z_{j_{o}})\cdot
 G_{t_c} \ket{\beta}\cdot && \non \\ 
\cdot \bra{\eta}
\check{G}_{t_c}^{-1}\cdot
\prod_{p_{i}}\Psi_{\delta_{p_{i}}}(x_{p_{i}}) 
\prod_{j_{i}}\Phi_{h_{j_{i}}}(z_{j_{i}},\bar z_{j_{i}})\cdot
\check{G}_{t_c} \ket{0} ~~~~~~~  &&
\end{eqnarray}
where $\mathbf{C}^{\beta }_{\alpha \; \eta}$ is the
structure constant of the operator algebra, and $\check{G}_{t_c}$ the
element of $Vir$ representing the map $\check{g}_{t_c}(z)$ defined on $D$.
In fact only the fields for which 
$$\sum_{\beta ,\eta}\mathbf{C}^{\beta }_{\alpha \; \eta}
(\varepsilon^{(0)}_t)^{\delta_{\beta}-\delta_{\alpha}-
 \delta_{\eta}}(\varepsilon^{(1)}_t)^{\delta_{\eta}}$$
is dominant are really significant in the above sum. Note that from the
conformal field theory viewpoint the scaling $\varepsilon^{(1)}_t \sim
(\varepsilon^{(0)}_t)^2$ (which is also suggested by independent
arguments for smooth $\gamma_t$) yields a formula for which the domain
$D$ and its complement appear symmetrically. We do not know if this
scaling relation holds for a typical $SLE_{\kappa}$ trace 
($\kappa \in ]4,8[$), but when subdomains of $\mathbf{H}$ containing
field insertions are swallowed by the $SLE_\kappa$ hull,
full conformal field theory correlation functions survive the swallowing.

\vskip 1.0 truecm

\section{Conclusions and conjectural perspectives.}
We have established a precise connection between $SLE_\kappa$
evolutions and boundary conformal field theories, with central charge
$c_\kappa$, eq.(\ref{dingdong}). Geometrically, this relation is based
on identifying the hull boundary state $\ket{K_t}$, eq.(\ref{kate}),
which encodes the evolution of the $SLE_\kappa$ hull.  The key point
is that this state, which belongs to a Virasoro module possessing a
null vector at level two, is conserved in mean. It is thus a
generating function for $SLE_\kappa$ martingales. The simplest
illustration of this property is provided by the algebraic derivation
of generalized crossing probabilities that we gave in Section 5. It
yields solutions of $SLE_\kappa$ stopping time problems in terms of
CFT correlation functions. It is clear that the method we described
can be generalized to deal with many more examples involving
multipoint correlation functions than just the ones we presented in
this paper.

Although we point out a direct relation between $SLE_\kappa$ evolution 
and 2D gravity via the KPZ formula (\ref{kpz00}), this is
clearly calling for developments, in order in particular to make
contact with ref.\cite{bertrand}.

Knowing the relation between crossing probabilities -- or more
generally stopping time problems -- and conformal
correlation functions, it is natural to wonder whether conformal field
theories can be reconstructed from $SLE_\kappa$ data. In view of
the explicit examples, eg
eqs.(\ref{deuxpoints},\ref{excur0},\ref{unpoint}),
and their multipoint generalizations,
it is tempting to conjecture that appropriate choices for stopping
time $\tau$ and state $\bra{v^*}$ in the relation
$${\bf E}[\bra{v^*}\, G_\tau\ket{\omega}]=\langle{v^*}\ket{\omega}$$ 
should lead to a reconstruction formula for, say, the
boundary fields, expressing $$\bra{0} \prod_j \Psi_{\delta_j}(x_j)
\ket{\omega}$$ as a linear combination of $SLE_\kappa$ expectations of the
form
$$
{\bf E}[ {\bf 1}_{r(\tau_1,\cdots ,\tau_n)}\,
C(x_1,\cdots, x_n)_{\delta_1,\cdots ,\delta_n} ] 
$$
where $r(\tau_1,\cdots ,\tau_n)$ specifies a relation between the
swallowing times of the points $x_1,\cdots,x_n$ and $C(x_1,\cdots,x_n)_{\delta_1,\cdots
\delta_n}$ is an appropriate function depending locally on the 
$f_{\tau_j}(x_j)$ as in
eq.(\ref{deuxpoints},\ref{excur0}). 
The analogue ansatz for bulk operators
would also involve the events coding for the position of the
$SLE_\kappa$ trace with respect to the insertion points of the bulk
operators as in eq.(\ref{unpoint}) and its multipoint generalizations.
We plan to report on this problem in a near future \cite{aplus}.

The existence of such reconstruction scheme would indicate that
$SLE_\kappa$ evolutions, or its avatars, provide alternative
formulations of, at least conformal, statistical field theories.
Contrary to usual approaches, the peculiarity of this, yet virtual,
reformulation of field theories would be that its elementary objects
are non-local and extended, with, at an even more hypothetical level, 
possible applications to (a dual formulation of) gauge theories.

\vskip 2.0 truecm

\section{Appendix A: Deterministic Loewner evolutions.}
\label{sec:dle}

The simplest example of deterministic Loewner evolution is when the
 driving term vanishes, so that the equation reduces to $\dot{g}=2/g$,
 which combined with $g_{t=0}=z$ leads to $g^2=z^2+4t$. When $z$ is
 real, $g$ is real as well, but $g$ is real also when $z$ is pure
 imaginary, $z=2i\sqrt{t'}$, $t' \in ]0,t]$. So the hull is
 $K_t=\{2i\sqrt{t'},t' \in [0,t]\}$, and with a slight abuse of
 notation, we write $g=\sqrt{z^2+4t}$, being understood that the
 determination of the square root is choosen to make $g$ continuous on
 $\overline{\bf H}\setminus K_t$ and behave like $z$ at $\infty$.

\vspace{0.5cm}

This trivial example allows to construct a more instructive one : by
appropriate fractional linear transformations in the source and image of $g(z)$, one can
construct a conformal representation for the situation when the hull
is an arc of circle of radius $r$ centered at the origin and emerging
from the real axis at $r$. Straightforward computations lead to a
family of maps
$$ g_{\lambda}(z):=-r
\frac{\frac{2\lambda+1}{\lambda+1}\sqrt{\lambda+\left(\frac{z-r}{z+r}\right)^2}
-\frac{2\lambda-1}{\sqrt{\lambda+1}}}
{\sqrt{\lambda+\left(\frac{z-r}{z+r}\right)^2}-\sqrt{\lambda+1}}
$$

The determination of $\sqrt{\lambda+\left(\frac{z-r}{z+r}\right)^2}$
is chosen in such a way that the behavior at large $z$ is
$g_{\lambda}(z)=z+O(1/z)$.  The parameter $r$ simply sets the scale, but the
nonnegative parameter $\lambda$ sets the angular extension $\theta$ of
the arc : apart from the real axis, the points $z$ such that $g_{\lambda}(z)$ is
real are of the form $z=re^{i\vartheta}, \vartheta \in [0,\theta]$
with $\tan^2 \theta/2 = \lambda$. 

When $\lambda \rightarrow +\infty$,
$\theta \rightarrow \pi$, and the free end of the arc approches the
real axis at the point $-r$. Our aim is to observe what happens in
this limit, especially to the points that are being ``swallowed'' (i.e.
the points inside the open half disc $D_r$ of radius $r$ centered at
the origin).

The determination of $\sqrt{\lambda+\left(\frac{z-r}{z+r}\right)^2}$
is such that on the two sides of the cut it is equal to 
$\pm \sqrt{\tan^2 \theta/2-\tan^2
  \vartheta/2}$ for $z=r^\pm e^{i\vartheta}$.
Explicit computation shows that 
$$
g_{\lambda}(-r)=-r\frac{2\lambda+1}{\lambda+1} \quad, \quad
g_{\lambda}(re^{i\theta})=-r\frac{2\lambda-1}{\lambda+1}$$
and $$
g_{\lambda}(r^\pm)=\pm
2r\sqrt{\frac{\lambda}{\lambda+1}}+\frac{r}{\lambda+1}.
$$

Moreover, $\sqrt{\lambda+\left(\frac{z-r}{z+r}\right)^2}$ is pure
imaginary when $z=re^{i\vartheta}, \vartheta \in [\theta,\pi]$. As
$g_{\lambda}$ is a real homographic function of
$\sqrt{\lambda+\left(\frac{z-r}{z+r}\right)^2}$, we conclude that
$\{re^{i\vartheta}, \vartheta \in [\theta,\pi]\}$ is mapped to a
semicircle by $g_{\lambda}$, and $D_r$ is mapped to the open half disc
bounded by this semicircle and a segment of the real axis.  

The important observation is that for large $\lambda$, $g_{\lambda}(r^+)
\rightarrow 2r$, while $g_{\lambda}(-r)$, $g_{\lambda}(r^-)$ and
$g_{\lambda}(re^{i\theta})$ have the same limit $-2r$, so that the
image of $D_r$ shrinks to a point.
On the other hand, if $z \in {\bf H} \setminus\overline{D}_r$, $\lim_{\lambda
\rightarrow +\infty} g_{\lambda}(z)=z+r^2/z$, so that the image of ${\bf H}
\setminus\overline{D}_r$ is ${\bf H}$: 
$$
\lim_{\lambda\to\infty}g_\lambda(z)=\cases{ z+ r^2/z&, for $|z|\geq r$ \cr
                                        -2r &, for $|z|<r$ \cr}
$$
The approach to the limit is interesting. Let 
$g_\lambda(z)=-2r+x_\lambda(z)+iy_\lambda(z)$ with $y_\lambda(z)>0$ by
construction. For $z\in D_r$, we have $x_\lambda(z)=2r/\lambda 
+O(1/\lambda^2)$ and $y_\lambda(z)=O(1/\lambda^2)$ so that
$\lim_{\lambda\to\infty}x_\lambda(z)/y_\lambda(z)=+\infty$ 
as expected for a loop
surrounding the point $z$ positively.

The rate at which the points in $D_r$ go to $-2r$ is also interesting.
If $z \in D_r$ one checks that, for large $\lambda$,
$g_{\lambda}(z)-g_{\lambda}(re^{i\theta})=-r/\lambda+O(1/\lambda^2)$,
so that if we take another $z'\in D_r $, the ratio
$$\frac{g_{\lambda}(z)-g_{\lambda}(re^{i\theta})}
{g_{\lambda}(z')-g_{\lambda}(re^{i\theta})}\rightarrow 1,$$
showing
that the points $z$ and $z'$ are come close to each other faster than
they approach the collapse point.

Going one step further in the expansion gives 
$$g_{\lambda}(z)-g_{\lambda}(re^{i\theta})=-\frac{r}{\lambda}+\frac{r}{\lambda^2}
\left(1+\frac{rz}{(r+z)^2}\right)+O(1/\lambda^3),$$
and one checks
that the map that appears at order $\lambda^{-2}$ maps $D_r$
conformally onto ${\bf H}$.

Up to now, we have not given a description of this example in terms of
Loewner evolution.  Explicit computation shows that
$$\frac{\partial g_{\lambda}(z)}{\partial\lambda}= \frac{2r^2}{(\lambda
+1)^3}\ \frac{1}{g_{\lambda}(z)+r\frac{2\lambda-1}{\lambda+1}}.$$
So, to deal with a normalized Loewner evolution,  we need to make a
change of evolution parameter. We set 
$$t\equiv \frac{r^2}{2}\left( 1-\frac{1}{(\lambda+1)^2}\right),\quad
\xi _t \equiv -r\frac{2\lambda-1}{\lambda+1},$$
and, with an abuse of notation, write $g_{t}$ for $g_{\lambda (t)}$.
Then $\partial_t{g_{t}}(z)=2/(g_{t}(z)-\xi _t)$ as usual.
The semicircle closes at $t_c=r^2/2$. For $z \in D_r$ and $t$
close to $t_c$, 
$$g_{t}(z)-\xi_t=
-(2(t_c-t))^{1/2}+2(t_c-t)\frac{z}{(r+z)^2}+O((t_c-t)^{3/2}.$$

\section{Appendix B: Virasoro intertwiners and their correlation
  functions.}
Here, we gather a few basic informations on Virasoro intertwiners.

Let us first deal with boundary operators $\Psi_\delta(x)$,
acting  from one highest weight Virasoro module $\CV_{r}$ to another 
module $\CV_{l}$ with respective highest weight vectors $\ket{h_r}$ and
$\ket{h_l}$:
$$
 \Psi_\delta(x):\ \CV_{r} \longrightarrow \CV_{l}
$$
They are constrained by the intertwinning relations (\ref{inter1}).
When $\CV_{r}$ and  $\CV_{l}$ are irreducible Verma modules, the space
of intertwinners between them is one dimensional. When $\CV_{l}$ is an
irreducible module, this space is $0$ or $1$ dimensional. This is so
because in these cases, all the matrix elements of $\Psi_\delta(x)$
can be obtained from the scalar products
$$
\bra{h_l}\prod_pL_{n_p}\Psi_\delta(x)\prod_qL_{-n_q}\ket{h_r}
$$
and these are fixed by the three-point function
$\bra{h_l}\Psi_\delta(x)\ket{h_r}$ and the intertwinning relations 
(\ref{inter1}).

Commutation relation with $L_0$ fix the scaling form of the three
point function:
$$
\bra{h_l}\Psi_\delta(x)\ket{h_r}= \mathbf{C}_{\delta\, h_r}^{h_l}\
x^{h_l-h_r-\delta} 
$$
The constant $\mathbf{C}_{\delta\, h_l}^{h_r}$ is called the structure
constant. The intertwiner $\Psi_\delta(x)$ 
from $\CV_r$ to $\CV_l$ exists whenever this
constant does not vanish. The fusion rules are those imposed by
demanding that $\mathbf{C}_{\delta\, h_r}^{h_l}$ be non zero.

If $\delta$ is generic and $\CV_l$ and $\CV_r$ are the possibly
reducible Virasoro Verma modules no constraint is imposed on the
possible values of the weight $h_l$, $h_r$ and $\delta$.

Constraints on $\mathbf{C}_{\delta\, h_r}^{h_l}$ arise when the Virasoro
modules possess null vector. As we shall only deal with null vectors
at level two, let us concentrate on the case where $\CV_r=\CH_{1,2}$ 
with $\ket{h_r}=\ket{\om}$ and $\delta$ is generic. 
Recall that $\ket{\om}$ has weight $h_{1,2}^\kappa=(6-\kappa)/2\kappa$.
Hence, $\Psi_\delta(x):\CH_{1,2} \longrightarrow \CV_{l}$. Since
$\ket{n_{1,2}}=(-2L_{-2}+\frac{\kappa}{2}L_{-1}^2)\ket{\omega}$
vanishes in $\CH_{1,2}$ we have:
$$
\bra{h_l}\Psi_\delta(x)\,
(-2L_{-2}+\frac{\kappa}{2}L_{-1}^2)\ket{\omega}=0 
$$
By using the intertwining relation (\ref{inter1}) to move the $L_{-n}$
to the left, this translates into
$$
\Big(\, 2\ell_{-2}^\delta(x) 
+\frac{\kappa}{2} \ell_{-1}^{\delta}(x)^2\,\Big)
\bra{h_l}\Psi_\delta(x)\,\ket{\om}=0
$$
This either imposes to $\mathbf{C}_{\delta \om}^{h_l}$ to vanish,
or $h_l-\delta-h_{1,2}^\kappa$ to be one of the two solutions
$\Delta_\pm(\delta)$ of eq.(\ref{kpz00}). In other words, the only
Virasoro intertwiners acting on $\CH_{1,2}$ are
$$
\Psi_\delta(x)\ :\ \CH_{1,2} \longrightarrow \CV_{h_\pm(\delta)},\quad
h_\pm(\delta)=\delta+h_{1,2}^\kappa+\Delta_\pm(\delta)
$$
with $\Delta_\pm(\delta)$ given in eq.(\ref{kpzsol}).
These are the fusion rules. For  instance, in the particular case
$\delta=h_{1,2}^\kappa$ then $h_\pm(h_{1,2}^\kappa)$ takes the two
possible values:
$$
h_\pm(h_{1,2}^\kappa) \in \{ 0 \ ;\ 
h_{1,3}^\kappa=\frac{8-\kappa}{\kappa}\}
$$
Symmetrically, the only interwiners $\Psi_\delta(x):\CH_{1,2}\to
\CV_{h_{1,2}^\kappa}$ which couple $\ket{\omega}$ to the state
$\ket{h_{1,2}^\kappa}$ with identical conformal weights should have
$\delta=0$ or $\delta=h_{1,3}^\kappa$, since then
$\Delta_\pm(\delta)+\delta=0$.

Multipoint correlators are computed by composing intertwiners.
For instance the four point function
$\bra{h}\Psi_{\delta_1}(x_1)\Psi_{\delta_2}(x_2)\ket{\omega}$ may be
thought as a matrix elements of the composition of two intertwiners:
$$
\Psi_{\delta_1}(x_1)\circ\Psi_{\delta_2}(x_2):
\CH_{1,2}\longrightarrow \CV_{h_\pm(\delta_2)} \longrightarrow \CV_h
$$
The fact that there are two possible choices for $h_\pm(\delta_2)$
is what is meant by the fact that there are two possible conformal
blocks.

The differential equations satisfied by the four point functions
also follow from the existence of the null vector $\ket{n_{1,2}}$
since: 
$$
\bra{h}\Psi_{\delta_1}(x_1)\Psi_{\delta_2}(x_2)
(-2L_{-2}+\frac{\kappa}{2}L_{-1}^2)\ket{\omega}=0 
$$
Using again the intertwining relations (\ref{inter1}) to move the
Virasoro generators to the left gives: 
\debut 
\Big[\,
2(\ell_{-2}^{\delta_1}(x_1) +\ell_{-2}^{\delta_2}(x_2))
+\frac{\kappa}{2} (\ell_{-1}^{\delta_1}(x_1)+\ell_{-1}^{\delta_2}
(x_2))^2\,\Big]
\bra{h}\Psi_{\delta_1}(x_1)\Psi_{\delta_2}(x_2)\ket{\omega}=0 \non
\fin
These are partial differential equations which turn into second order
ordinary differential equations, once the trivial
scaling behaviors have been factorized. Some examples appear in the
main text.

\vskip .5 truecm

Consider now bulk intertwiners $\Phi_h(z,\bar z)$ of dimension $h$. 
The logic is the same as before except that the
intertwining relations (\ref{inter2}) are slightly more involved.
In particular, $\Phi_h(z,\bar z)$ acting from one module to another
one
$$
\Phi_h(z,\bar z): \CV_r\longrightarrow \CV_l
$$
is also determined by its three point function 
$\bra{h_l}\Phi_h(z,\bar z)\ket{h_r}$ although, contrary to
boundary operators, the intertwining relations (\ref{inter2}) do not
completely fix it. $\bra{h_l}\Phi_h(z,\bar z)\ket{h_r}$ is
generally more involved than simply a power law.

As before no constraint is imposed if $h$ is generic and $\CV_l$ and
$\CV_r$ are Verma modules. Constraints arise if one of the two modules
possesses null vectors. For simplicity, assume that $\CV_r=\CH_{1,2}$
and $\ket{h_r}=\ket{\om}$ as almost surely everywhere in this paper,
so that:
 $$
\Phi_h(z,\bar z): \CH_{1,2}\longrightarrow \CV_l
$$
Then, since $\ket{n_{1,2}}$ vanishes in $\CH_{1,2}$, 
$$
\bra{h_l}\Phi_h(z,\bar z)
(-2L_{-2}+\frac{\kappa}{2}L_{-1}^2)\ket{\omega}=0 
$$
or equivalently,
\debut
\Big[\, 2(\ell_{-2}^{h}(z) +\ell_{-2}^{h}(\bar z))
+\frac{\kappa}{2} (\ell_{-1}^{h}(z)+\ell_{-1}^{h}(\bar z))^2\,\Big]
\bra{h_l}\Phi_h(z,\bar z)\ket{\omega}=0
\label{diffbulk}
\fin
Generically this equation has two independent solutions implying that the space
of Virasoro intertwiners acting in $\CH_{1,2}$ is two dimensional.
This is what is meant by the fact that there are two conformal blocks
in the bulk.

Further constraints, implying fusion rules,
 arise if we demand that the image space $\CV_l$
also possesses null vectors.
Bulk operators may be thought of as the compositions of two chiral
intertwiners depending respectively on $z$ and $\bar z$.

\vskip 0.5 truecm

Finally, besides the highest weight Verma modules ${\cal V}_h$ one may 
also consider the contravariant representations $\tilde {\cal V}_h$
induced on their dual spaces. For generic $h$, ${\cal V}_h$ and
$\tilde {\cal V}_h$ are isomorphic as Virasoro modules, but they are
not if ${\cal V}_h$ possesses sub-modules. In particular 
${\cal V}_{h_{1,2}^\kappa}$ and $\tilde {\cal V}_{h_{1,2}^\kappa}$
are not isomorphic but $\tilde {\cal V}_{h_{1,2}^\kappa}$ contains a
submodule isomorphic to ${\cal H}_{1,2}$. Contravariant modules
arise when defining conformal fields. Namely, one may
show that the intertwinners $\Psi_\delta(x)$,
$$
\Psi_\delta(x): {\cal V}_h \longrightarrow \tilde 
{\cal V}_{h_{1,2}^\kappa}
$$ from the Verma module ${\cal V}_h$ to the contravariant module 
$\tilde {\cal V}_{h_{1,2}^\kappa}$ exist, while there are no
intertwinner from ${\cal V}_h$ to ${\cal V}_{h_{1,2}^\kappa}$,
for generic $\delta$.
If the fusion rules are satisfied, then $\Psi_\delta(x)$ maps
${\cal V}_h$ into ${\cal H}_{1,2}$, viewed as 
$\tilde {\cal V}_{h_{1,2}^\kappa}$ submodules.
This is the subtilities alluded in footnote 6, eq.(\ref{Fdeux}).

\section{Appendix C: Derivation of formula (\ref{eq:grosse}).}

To derive eq.\ref{eq:grosse}, we manipulate 
$$ \bra{0} 
\prod_{p_{o}}\Psi_{\delta_{p_{o}}}(x_{p_{o}})
\prod_{j_{o}}\Phi_{h_{j_{o}}}(z_{j_{o}},\bar z_{j_{o}})\cdot
\prod_{p_{i}}\Psi_{\delta_{p_{i}}}(x_{p_{i}})
\prod_{j_{i}}\Phi_{h_{j_{i}}}(z_{j_{i}},\bar z_{j_{i}})\cdot
G_t \ket{\alpha}.$$
Remember that the ``$i(n)$'' arguments are inside $D$ or
$I$, 
and the ``$o(ut)$'' arguments sit outside.

We conjugate by $G_t$ and then insert the identity  operator
$\mathbf{1}=\sum_{\beta}\ket{\beta}\bra{\beta}$ to separate the
``$i(n)$'' and the ``$o(ut)$'' contributions, leading to:
\begin{eqnarray*}
\sum_{\beta} &  \bra{0}{
  \prod_{p_{o}}[\mony_t'(x_{p_{o}})]^{\delta_{p_{o}}}
  \Psi_{\delta_{p_{o}}}(\mony_t(x_{p_{o}})) \cdot \prod_{j_{o}}
  |\mony_t'(z_{j_{o}})|^{2h_{j_{o}}}
  \Phi_{h_{j_{o}}}(\mony_t(z_{j_{o}}),\bar \mony_t(\bar
  z_{j_{o}}))} \ket{\beta}  & \\ & \bra{\beta}{
  \prod_{p_{i}}[\mony_t'(x_{p_{i}})]^{\delta_{p_{i}}}
  \Psi_{\delta_{p_{i}}}(\mony_t(x_{p_{i}})) \cdot \prod_{j_{i}}
  |\mony_t'(z_{j_{i}})|^{2h_{j_{i}}}
  \Phi_{h_{j_{i}}}(\mony_t(z_{j_{i}}),\bar \mony_t(\bar
  z_{j_{i}}))} \ket{\alpha}. & 
\end{eqnarray*}

The ``$o(ut)$'' part behaves nicely when $t \nearrow t_c$, but for the
``$i(n)$'' part we write $\mony_t(z)=\varepsilon^{(0)}_t+
\varepsilon^{(1)}_t\check{\mony}_{t}(z)$ and use the covariance under
similarities : 
\begin{eqnarray*}
&&\bra{\beta}
  \prod_{p_{i}}[\mony_t'(x_{p_{i}})]^{\delta_{p_{i}}}
  \Psi_{\delta_{p_{i}}}(\mony_t(x_{p_{i}})) \cdot \prod_{j_{i}}
  |\mony_t'(z_{j_{i}})|^{2h_{j_{i}}}
  \Phi_{h_{j_{i}}}(\mony_t(z_{j_{i}}),\bar  \mony_t(\bar
  z_{j_{i}}))  \ket{\alpha} = \\
&& ~~~~~~~~~~ (\varepsilon^{(1)}_t)^{\delta_{\beta}-\delta_{\alpha}}\
\bra{\beta}\Psi_{\delta_{\alpha}}\left(-\varepsilon^{(0)}_t/
\varepsilon^{(1)}_t\right) \prod_{p_{i}} 
[\check{\mony}_t'(x_{p_{i}})]^{\delta_{p_{i}}}
  \Psi_{\delta_{p_{i}}}(\check{\mony}_t(x_{p_{i}})) \cdot \\
&& ~~~~~~~~~~~~~~~~\cdot \prod_{j_{i}}|\check{\mony}_t'(z_{j_{i}})|
^{2h_{j_{i}} 
\Phi_{h_{j_{i}}}(\check{\mony}_t(z_{j_{i}}),\bar{\check{\mony}}_t(\bar
z_{j_{i}}))} \ket{0} 
\end{eqnarray*}
where 
$\Psi_{\delta_{\alpha}}$ is the field such that
$\Psi_{\delta_{\alpha}}(0)\ket{0}=\ket{\alpha}$. Upon insertion of the
identity operator $\mathbf{1}=\sum_{\eta}\ket{\eta}\bra{\eta}$ to
separate $\Psi_{\delta_{\alpha}}$ from the ``$i(n)$'' fields, we find
\begin{eqnarray*}
&& \bra{\beta}{ \prod_{p_{i}}[\mony_t'(x_{p_{i}})]^{\delta_{p_{i}}}
  \Psi_{\delta_{p_{i}}}(\mony_t(x_{p_{i}})) \cdot \prod_{j_{i}}
  |\mony_t'(z_{j_{i}})|^{2h_{j_{i}}}
  \Phi_{h_{j_{i}}}(\mony_t(z_{j_{i}}),\bar \mony_t(\bar z_{j_{i}}))}
\ket{\alpha} =\\
&& ~~~~~~~~~~~ \sum_{\eta} (\varepsilon^{(1)}_t)^{\delta_{\beta}
-\delta_{\alpha}}
\left(\frac{\varepsilon^{(0)}_t}{
\varepsilon^{(1)}_t}\right)^{\delta_{\beta}-\delta_{\alpha}-\delta_{\eta}} 
\bra{\beta}\Psi_{\delta_{\alpha}}(1)\ket{\eta}\cdot\\
&& \cdot \bra{\eta}{\prod_{p_{i}}  
[\check{\mony}_t'(x_{p_{i}})]^{\delta_{p_{i}}}
  \Psi_{\delta_{p_{i}}}(\check{\mony}_t(x_{p_{i}})) \cdot \prod_{j_{i}}
  |\check{\mony}_t'(z_{j_{i}})|^{2h_{j_{i}}}
  \Phi_{h_{j_{i}}}(\check{\mony}_t(z_{j_{i}}),\bar{\check{\mony}}_t(\bar
  z_{j_{i}}))} \ket{0} 
\end{eqnarray*}
In fact $\bra{\beta}\Psi_{\delta_{\alpha}}(1)
\ket{\eta} = \mathbf{C}^{\beta }_{\alpha \; \eta}$ is the 
structure constant of the operator algebra. When $t \nearrow t_c$,
$\check{\mony}_t$ has a limit which maps conformally $D$ onto the
upper half plane, and we may represent it by an operator
$\check{G}_{t_c}$.
This leads to formula (\ref{eq:grosse}).


\end{document}